\documentclass[showpacs,preprintnumbers,amsmath,reprint,aps,prx,superscriptaddress]{revtex4-1}

\usepackage{bm}
\usepackage{graphicx}
\usepackage{color}
\usepackage{amsmath}
\usepackage{hyperref}
\usepackage{epstopdf}
\usepackage{upgreek}
\usepackage{mathptmx, textcomp}
\usepackage[latin1]{inputenc}
\usepackage[T1]{fontenc}
\usepackage{array, multirow}
\usepackage[table]{xcolor}
\usepackage{booktabs}
\usepackage{longtable}

\usepackage[normalem]{ulem}

\newcommand{\commentOut}[1]{}

\newcommand{\ket}[1]{\left|#1\right>}
\newcommand{\bra}[1]{\left<#1\right|}

\begin{document}

\title{Observation of spin-orbit-dependent electron scattering using long-range Rydberg molecules}

\author{Markus Dei{\ss}}
\affiliation{Institut f\"{u}r Quantenmaterie and Center for Integrated Quantum Science and
	Technology IQ$^{ST}$, Universit\"{a}t Ulm, 89069 Ulm, Germany}
\author{Shinsuke Haze}
\affiliation{Institut f\"{u}r Quantenmaterie and Center for Integrated Quantum Science and
	Technology IQ$^{ST}$, Universit\"{a}t Ulm, 89069 Ulm, Germany}
\author{Joschka Wolf}
\affiliation{Institut f\"{u}r Quantenmaterie and Center for Integrated Quantum Science and
	Technology IQ$^{ST}$, Universit\"{a}t Ulm, 89069 Ulm, Germany}
\author{Limei Wang}
\affiliation{Institut f\"{u}r Quantenmaterie and Center for Integrated Quantum Science and
	Technology IQ$^{ST}$, Universit\"{a}t Ulm, 89069 Ulm, Germany}
\author{Florian Meinert}
\affiliation{5. Physikalisches Institut and Center for Integrated Quantum Science and
	Technology IQ$^{ST}$, Universit\"{a}t Stuttgart, 70569 Stuttgart, Germany}
\author{Christian Fey}
\affiliation{Zentrum f\"{u}r optische Quantentechnologien, Universit\"{a}t Hamburg, Fachbereich Physik, 22761 Hamburg, Germany}
\author{Frederic Hummel}
\affiliation{Zentrum f\"{u}r optische Quantentechnologien, Universit\"{a}t Hamburg, Fachbereich Physik, 22761 Hamburg, Germany}
\author{Peter Schmelcher}
\affiliation{Zentrum f\"{u}r optische Quantentechnologien, Universit\"{a}t Hamburg, Fachbereich Physik, 22761 Hamburg, Germany}
\affiliation{The Hamburg Centre for Ultrafast Imaging, Universit\"{a}t Hamburg, 22761 Hamburg, Germany}
\author{Johannes Hecker Denschlag}
\affiliation{Institut f\"{u}r Quantenmaterie and Center for Integrated Quantum Science and
	Technology IQ$^{ST}$, Universit\"{a}t Ulm, 89069 Ulm, Germany}

\date{\today}

\begin{abstract}	
We present experimental evidence for spin-orbit interaction of an electron as it scatters from a neutral atom. The scattering process takes place within a Rb$_2$ ultralong-range Rydberg molecule, consisting of a Rydberg atomic core, a Rydberg electron and a ground state atom. The spin-orbit interaction leads to characteristic level splittings of vibrational molecular lines which we directly observe via photoassociation spectroscopy. We benefit from the fact that molecular states dominated by resonant $p$-wave interaction are particularly sensitive to the spin-orbit interaction. Our work paves the way for studying novel spin dynamics in ultralong-range Rydberg molecules. Furthermore, it shows that the molecular setup can serve as a micro laboratory to perform precise scattering experiments in the low-energy regime of a few meV.
\end{abstract}

\maketitle
\section{Introduction}
\label{sec:intro}

Since their prediction almost twenty years ago \cite{Greene2000} and boosted by their first observation \cite{Bendkowsky2009}, ultralong-range Rydberg molecules have become a research area of major interest (for reviews, see, e.g. \cite{Shaffer2018, Fey2019, Eiles2019}). Nevertheless, the spin substructure of these molecules is not fully understood yet. In particular, one fundamental unresolved question concerns the coupling between the total electronic spin $\vec{S}$ and the relative orbital angular momentum $\vec{L}_\text{p}$ of the Rydberg electron with respect to the ground state perturber atom. The role of this $\vec{L}_\text{p} \cdot \vec{S}$ type spin-orbit interaction for the molecular system was predicted almost twenty years ago \cite{Khuskivadze2002}, and has remained a topic of active research until now \cite{Markson2016, Eiles2017}. From the experimental side, some preliminary indication for $\vec{L}_\text{p} \cdot \vec{S}$ coupling has been found recently \cite{Thomas2018}, however, clear evidence has been lacking. It has escaped discovery although a variety of spectroscopic studies with impressive resolution were carried out, investigating Rydberg molecules for various atomic species (Rb, Cs, Sr) and different Rydberg orbitals ($S$, $P$, or $D$) \cite{Bendkowsky2009, Bellos2013, Anderson2014a, Krupp2014, Kleinbach2017, Li2011, Tallant2012, Booth2015, Sassmannshausen2015, DeSalvo2015, Niederpruem2016a, Niederpruem2016b, Bendkowsky2010, Boettcher2016, MacLennan2018}.

Very recently, in parallel with our work reported here, {\it indirect} evidence for $\vec{L}_\text{p} \cdot \vec{S}$ coupling has been found in the observation of specific pendular states in Rb$_2$ ultralong-range Rydberg molecules \cite{Engel2019}, an effect predicted shortly before in \cite{Hummel2018}. Here, we complete the evidence for $\vec{L}_\text{p} \cdot \vec{S}$ interaction, as we spectroscopically \textit{directly} observe the Rydberg molecular level splitting caused by it; in fact, we resolve the full fine structure multiplet. For this, we investigate ultralong-range $^{87}$Rb$_2$ Rydberg molecules consisting of a $5S_{1/2}$ ground state atom and a $16P_{3/2}$ Rydberg atom. The molecular bound states of interest are located in the second outermost well of the Born-Oppenheimer potential energy curve (PEC), which is significantly influenced by the $p$-wave shape resonance. The resonant $p$-wave interactions strongly increase the $\vec{L}_\text{p} \cdot \vec{S}$-induced  level splittings of spin states so that they can be well resolved experimentally. For the molecular level spectroscopy, we carry out photoassociation in ultracold clouds of Rb ground state atoms, which can be prepared in different spin polarizations. We observe three vibrational ladders of the molecular Rydberg states. Each ladder has a characteristic line-multiplet substructure, which allows for unambiguous assignment of all spin states. Using model calculations on the basis of a pseudopotential Hamiltonian and including spin-spin and spin-orbit interactions \cite{Eiles2017}, we are able to fully explain the observed spectra.

\section{Molecular system and potential energy curves}
\label{sec:overview}

\begin{figure}[b]
	\includegraphics[width=0.9\columnwidth]{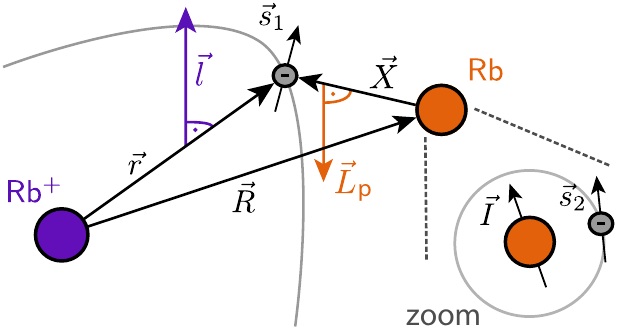}
	\caption{Composition of the molecular system (see text).}
	\label{fig1}
\end{figure}

\begin{figure*}[t]
	\includegraphics[width=2.05\columnwidth]{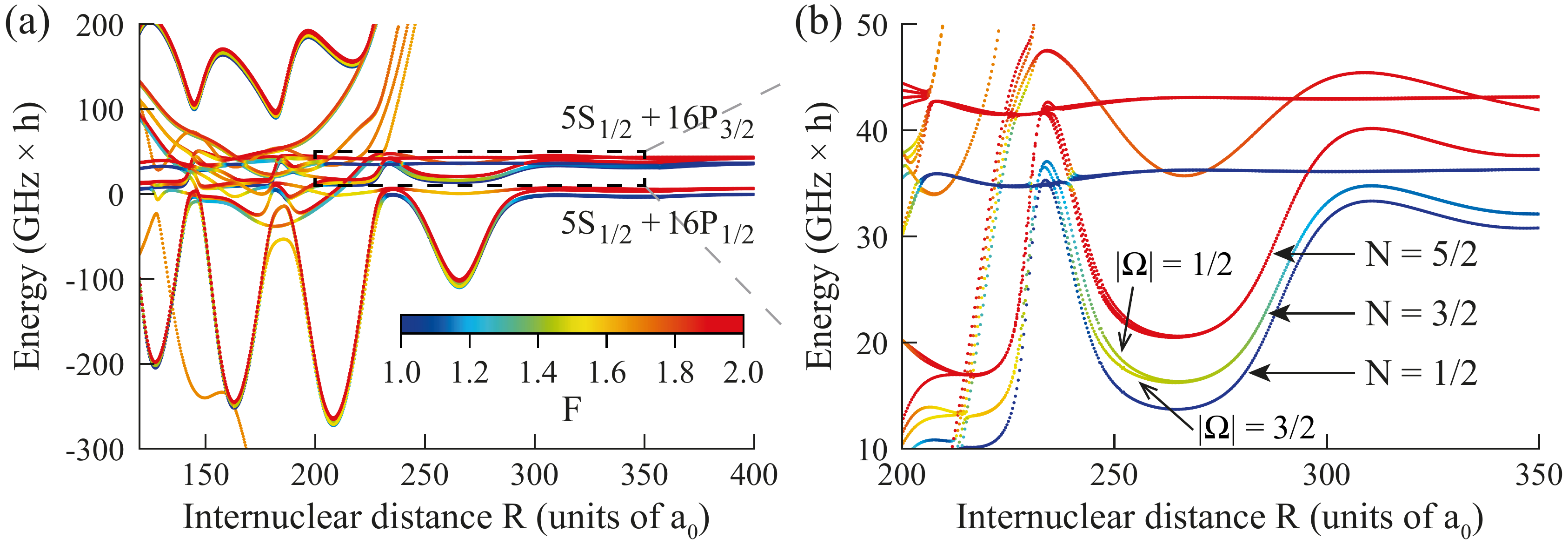}
	\caption{(a) The molecular PECs correlated to the $5S_{1/2}+16P_j$ atomic asymptotes for $j\in\{1/2,\,3/2\}$ and different hyperfine states $F\in \{1,\,2\}$ of the $5S_{1/2}$ atom. The color code represents the expectation value of the quantum number $F$. Calculations of PECs are described in Sec.$\:$\ref{subsec:model}. (b) Zoom into the dashed rectangle in (a) indicating the region of interest for the present work. $N$ and $|\Omega|$ are quantum numbers which label the PECs. The $N=5/2$, $3/2$, $1/2$ branches are composed of a triplet, doublet, and singlet substructure of $|\Omega|$ states, respectively.
	}
	\label{fig2}
\end{figure*}

The molecular system is sketched in Fig.$\:$\ref{fig1}. A ground state atom is located at position $\vec{R}$ relative to the ionic core of a Rydberg atom. The Rydberg electron at position $\vec{r}$ has spin $\vec{s}_1$ and orbital angular momentum $\vec{l}$ relative to the ionic core. Its total angular momentum is described by $\vec{j}=\vec{l}+\vec{s}_1$. The ground state atom possesses electronic spin $\vec{s}_2$ and nuclear spin $\vec{I}$ which are coupled by hyperfine interaction to form the total angular momentum $\vec{F}=\vec{I}+\vec{s}_2$ \cite{Anderson2014a, Anderson2014b, Sassmannshausen2015, Niederpruem2016b, Boettcher2016, MacLennan2018}. In the reference frame of the ground state atom the Rydberg electron is located at position $\vec{X}=\vec{r}-\vec{R}$ and has orbital angular momentum $\vec{L}_\text{p}$. Actually, we will be mainly interested in $\vec{L}_\text{p}\cdot \vec{S}$ spin-orbit coupling, where $\vec{S}$ is the total electronic spin $\vec{S} =\vec{s}_1 +\vec{s}_2$.

Figure \ref{fig2} shows the relevant PECs for our experiments. The ultralong-range Rydberg molecular states we investigate are bound in the second outermost wells at an internuclear distance of about $260\:a_0$. Here, $a_0$ is the Bohr radius. Figure \ref{fig2}(b) is a zoom onto these wells. On the left hand side of the wells steep butterfly PECs \cite{Chibisov2002,Hamilton2002,Niederpruem2016a} cross through that arise because of a $p$-wave shape resonance, where the Rydberg electron with angular momentum $L_\text{p} = 1$ resonantly interacts with the Rb ground state atom. This resonance occurs at a collision energy $E_r^\text{avg}=26.6\:\text{meV}$ \cite{noteEavg,Engel2019}. Due to the vicinity to the $p$-wave shape resonance the ultralong-range Rydberg molecular states in the second outermost wells experience strong $p$-wave interaction and are thus very sensitive to $\vec{L}_\text{p} \cdot \vec{S}$ coupling.

For large distances $R$, the $P$ state PECs have four asymptotes, corresponding to the combinations of the atomic Rydberg states $16P_{3/2}$ and $16P_{1/2}$ with hyperfine states $F = 1$ and $F = 2$ of the Rb ground state atom. The color coding in Fig.$\:$\ref{fig2} shows the $F$ content of the states (see also Fig.$\:$\ref{fig8} of the Appendix for the electronic spin $S$ content). As can be seen clearly in Fig.$\:$\ref{fig2}(b) some of the PECs exhibit $F$ mixing. As we will discuss in more detail in Sec.$\:$\ref{sec:calculations} this is especially due to the spin-dependence of the $p$-wave interaction.

In order to formally label the PECs, it is convenient to use the quantum number $N$ corresponding to the angular momentum $\vec N = \vec{S} + \vec{I}$. The spin-orbit interaction $\vec{L}_\text{p} \cdot \vec{S}$ splits up each PEC characterized by $N$, according to its multiplicity $N(N+1)$ into different $\Omega = -N, -N+1,\dots,N$ states. Here, $\Omega = m_I + m_S + m_l$ is the magnetic quantum number of the total angular momentum, and we have chosen the internuclear axis as the quantization axis. In Fig.$\:$\ref{fig2}(b) only the splitting of the $N = 3/2$ PEC into $|\Omega| = 1/2 $ and $|\Omega| = 3/2 $ is clearly visible. Because of rotational symmetry about the internuclear axis, the PECs for each pair of $\pm \Omega$ are generally energetically degenerate. Therefore, for each $N$, energy splittings only arise between the $(2N+1)/2$ different $|\Omega|$ components.

\section{Experiments and spectroscopic results}
\label{sec:expscheme}

The experiments are carried out in a hybrid atom-ion setup \cite{Schmid2012} consisting of a crossed optical dipole trap for an ultracold cloud of $^{87}$Rb ground state atoms and a linear Paul trap which we use in the detection of Rydberg molecules. The dipole trap operates at a wavelength of $1065\:\text{nm}$ and has a potential depth of about $20\:\upmu\text{K}\times k_\text{B}$.
The atomic sample is prepared either in the hyperfine state $F=1,m_F=-1$ or in the state $F=2,m_F=+2$. It has a temperature of $\approx 1\:\upmu\text{K}$, and typically consists of about $4\times 10^6$ atoms. The cloud is Gaussian-shaped with a size of $\sigma_{x,y,z} \approx (70,10,10)\:\upmu\text{m}$ along the three directions of space.

\begin{figure}[t]
	\includegraphics[width=\columnwidth]{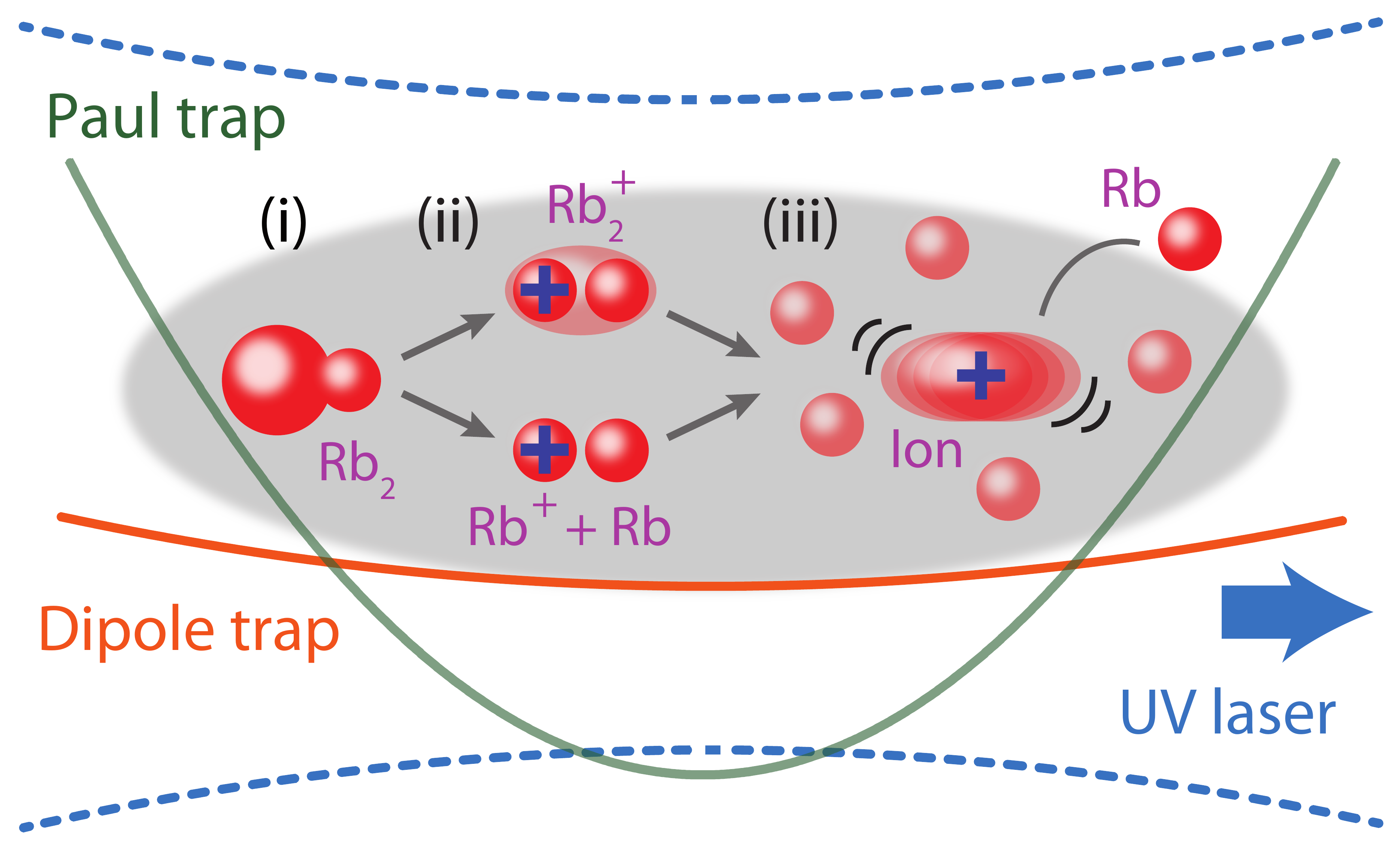}
	\caption{Illustration of the experimental setup and scheme.
		The orange solid line indicates the dipole trap potential for the ultracold neutral atoms while the dark-green solid line represents the Paul trap potential for ions. (i) Inside the atom cloud (indicated by the gray shaded area) Rb$_2$ Rydberg molecules are produced by means of the UV photoassociation laser (blue dashed lines and blue arrow). (ii) The molecules can decay into ions, where processes leading to Rb$^+$ and Rb$_2^+$ are possible (see, e.g., \cite{Niederpruem2015, Schlagmueller2016}). A resulting ion is captured by the Paul trap. (iii) The micromotion-driven ion elastically collides with Rb atoms leading to atom loss from the dipole trap.
	}
	\label{fig3}
\end{figure}

The general procedure of our experiment is as follows (see also the illustration in Fig.$\:$\ref{fig3}). We measure photoassociation spectra by scanning the frequency of a narrow-linewidth laser in a step-like fashion at a wavelength of about $302\:\text{nm}$ (for technical details on the photoassociation laser setup, see Sec.$\:$\ref{PA setup} of the Appendix). For each laser frequency we produce a cold cloud of Rb atoms and expose it for a well-defined time of typically a few hundred ms to the laser light. If the laser frequency is on resonance, photoassociation of $5S_{1/2}-16P_{3/2}$ Rb$_2$ Rydberg molecules takes place [see (i) Fig.$\:$\ref{fig3}]. We detect this production of dimers as follows. Because of various processes, such as photoionization, collisions, and ionization due to molecular relaxation,  some of the Rydberg molecules decay into ions (ii). These ions are subsequently confined in the linear Paul trap which has a trap depth of about $1\:\text{eV}$. The Paul trap is centered on the optical dipole trap so that the ions are immersed in the atom cloud. The ions inflict loss on the atom cloud (iii) \cite{Haerter2013b, Wolf2017}, which we measure via absorption imaging. Thus, by detecting atom loss, we infer the production of Rydberg molecules. In brief, the losses are due to micromotion-driven elastic collisions between atoms and ions, which expel atoms out of the shallow dipole trap. Even a single ion can lead to a significant loss signal. In general, the number of remaining atoms decreases with increasing number of ions.

\begin{figure}[t]
	\includegraphics[width=\columnwidth]{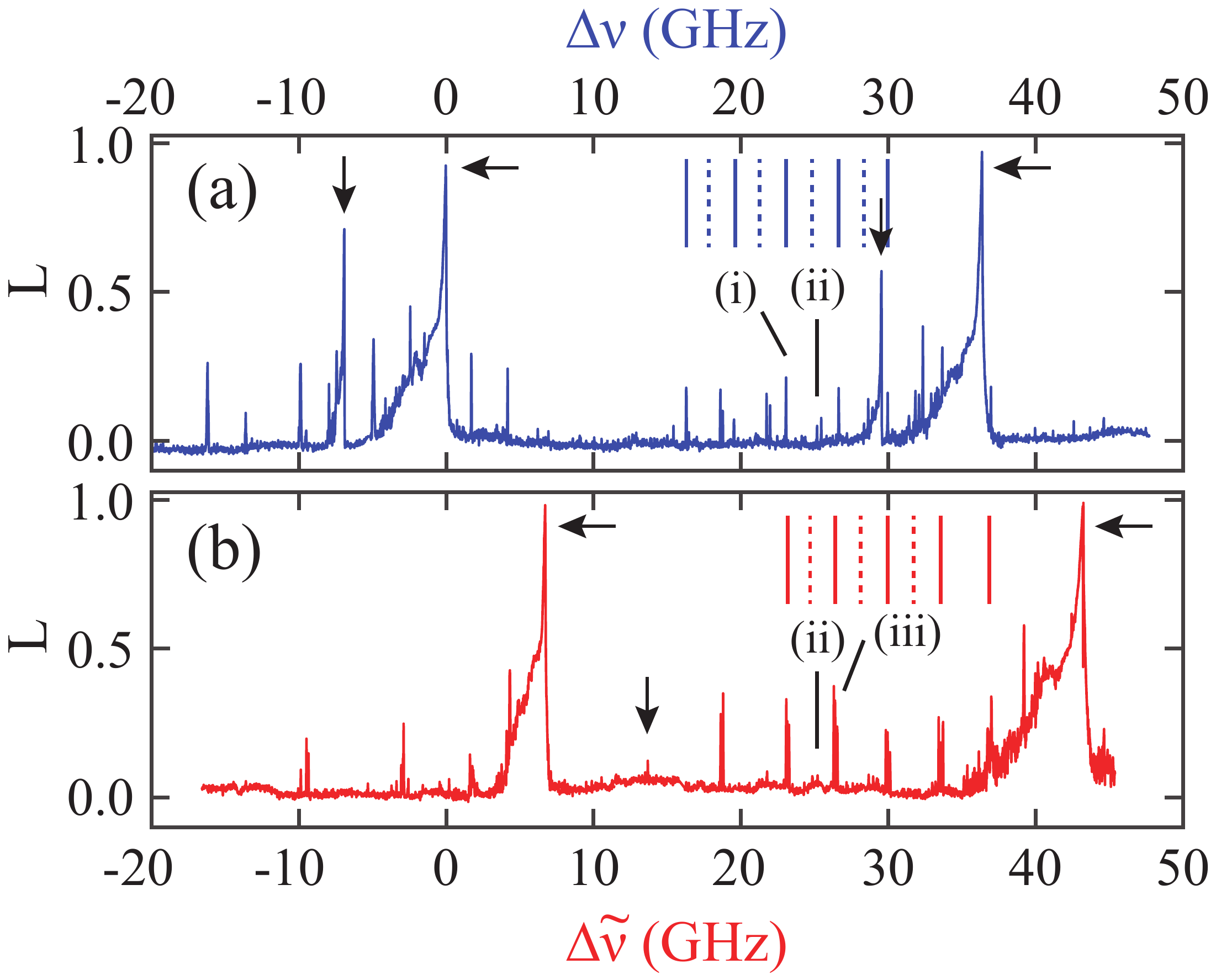}
	\caption{Spectra measured for atomic samples initially prepared in the hyperfine state $F=1$ (a) and $F=2$ (b), respectively. Shown is the atom loss $L$ as a function of the frequency $\nu$ of the UV spectroscopy laser light. The frequency $\nu$ is given in terms of $\Delta \nu=\nu-\nu_0$ (a) and $\Delta \tilde{\nu}=\nu-\nu_0+2\times \nu_\text{hfs}$ (b), where $\nu_0=991.55264\:\text{THz}$ is the resonance frequency for the $16P_{1/2}$ atomic Rydberg line when starting with $F=1$ atoms. The data of (a) are obtained for a pulse duration of $125\:\text{ms}$ of the spectroscopy light while for (b) $200\:\text{ms}$ are used (the light intensities, micromotion energies and ion-atom cloud interaction times are about the same for both scans). Horizontal and vertical black arrows mark resonances assigned to atomic transitions. The black solid lines with denotations (i), (ii), and (iii) point to line multiplets which are investigated in Fig.$\:$\ref{fig6} with higher resolution. Vertical red solid (dashed) lines in (a) illustrate the frequency positions of observed strong (weak) three line multiplets for $F=2$, while vertical blue solid (dashed) lines in (b) mark strong (weak) single line peaks for $F=1$.
	}
	\label{fig4}
\end{figure}

In Fig.$\:$\ref{fig4} two photoassociation spectra in the vicinity of the atomic $16P$ Rydberg state are presented. We plot the normalized atom loss $L=1-\tilde{N}/\tilde{N}_0$  as a function of the photoassociation laser frequency. Here, $\tilde{N}$ and $\tilde{N}_0$ are the remaining number of atoms after an experimental run when the photoassociation laser was turned on and off, respectively. The loss-signal strengths in Fig.$\:$\ref{fig4} have a strongly non-linear dependence on the number of trapped ions. While the largest loss signals correspond to hundreds of ions the smallest loss peaks are the result of only a few ions. For the measurements of Fig.$\:$\ref{fig4} the frequency $\nu$ of the photoassociation laser was scanned in steps of $20\:\text{MHz}$, and each data point represents a single run of the experiment. Scan (a) (blue data points) shows data for atoms prepared in the hyperfine state $F=1, m_F = -1$, while scan (b) (red data points) was obtained for atoms prepared in $F=2, \, m_F = +2$. For convenience, the two spectra are horizontally shifted relative to each other by twice the hyperfine splitting of the electronic ground state of $^{87}$Rb, i.e. $2 \times \nu_\text{hfs}=2 \times 6.835\:\text{GHz}$ \cite{Bize1999, Arimondo1977}, to account for the frequency spacing of the $F=1+F=1$ and $F=2+F=2$ atomic asymptotes there. Then, signals for identical molecular levels line up in both data sets of Fig.$\:$\ref{fig4}. Besides the photoassociation resonances the spectra also include the $16P_{3/2}$ and $16P_{1/2}$ atomic Rydberg lines which are marked with arrows. A discussion of the atomic lines is given in Sec.$\:$\ref{atomic lines} of the Appendix. In the following, we focus on the frequency range of $10\:\text{GHz}<\Delta \nu <40\:\text{GHz}$, where we expect our molecular Rydberg states of interest [c.f.  Fig.$\:$\ref{fig2}(b)].

\begin{figure*}[t]
	\includegraphics[width=2\columnwidth]{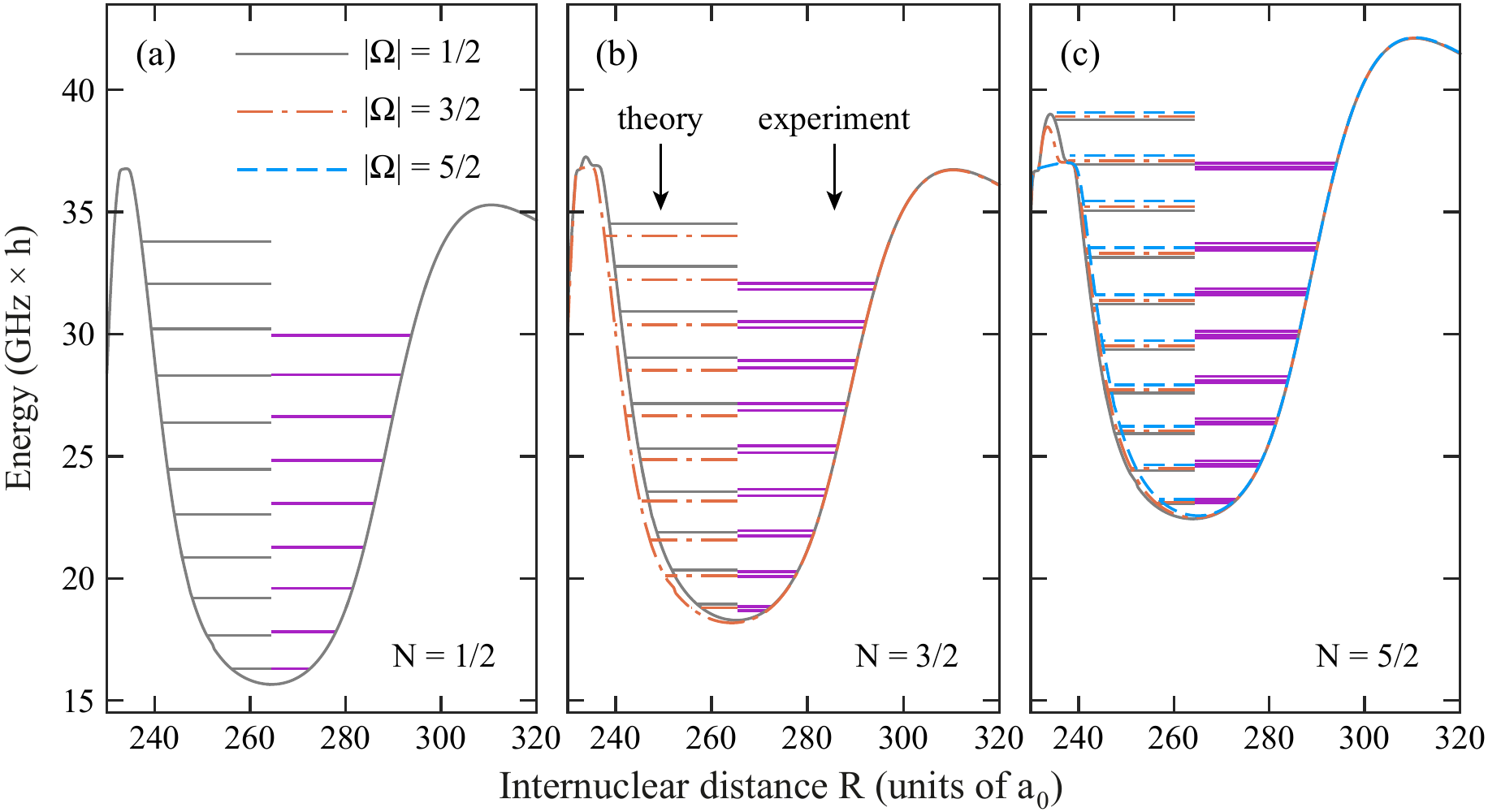}
	\caption{Comparison of measured molecular term energies (purple horizontal lines) and calculated molecular term energies for $N=1/2$ (a), $N=3/2$ (b), and $N=5/2$ (c). The theoretical results for the different $|\Omega|$ states are indicated by the line color and line style as given in the legend. Here, the predicted term energies and the PECs are shifted by $1.94\:\text{GHz}\times h$ to higher energies as compared to Table \ref{table1} of the Appendix and Fig.$\:$\ref{fig2}(b), respectively, for better comparison to the experimental data. Then, the calculated position for the vibrational ground state of $N=1/2$ (which has even symmetry) coincides with the lowest observed singlet line at $16.30\:\text{GHz}\times h$ (which is a strong signal). Such a shift is well within the uncertainty of a few $\text{GHz}\times h$ of absolute energy determinations in the perturbative electronic structure calculations \cite{Eiles2017, Fey2015}.
	}
	\label{fig5}
\end{figure*}

An analysis of our measured spectra shows that we observe three different vibrational ladders. The frequency spacings between vibrational lines for each ladder are approximately equidistant, typically ranging between 1.4 and 1.8$\:$GHz. The first of the three ladders appears in spectrum (a), the second ladder appears in spectrum (b), and lines of the third ladder appear in both spectra (a) and (b). The positions of signals of the first and second ladder are marked in Fig.$\:$\ref{fig4} as a progression of vertical solid and dashed lines, which indicate strong and weak transition lines, respectively. We note that not all of the experimentally observed lines are resolved in the two shown spectra (a) and (b). More refined scans over several small frequency ranges of interest revealed additional resonances (for more information on the methodology see Sec.$\:$\ref{Collection} in the Appendix). A list of all observed lines can be found in Table \ref{table1} of the Appendix. Signals of the third ladder are in general comparatively weak and some of these are barely or not visible in Fig.$\:$\ref{fig4}. The signal marked with (ii) in (a) and (b) is an example of a line from the third ladder. The selectivity of each vibrational ladder for being observed exclusively in the spectra (a) or (b) (or in both) can be explained by the total angular momentum $\vec{F}$ content of the PECs in Fig.$\:$\ref{fig2}(b). According to selection rules for electric dipole transitions, photoassociation does not intrinsically change the $F$ quantum number of the Rb atom that stays in the ground state. For example, when starting from an ensemble of $F = 1$ atoms only vibrational states in the potential wells with $N = 1/2$ and $N = 3/2$ can be reached because they exhibit some $F = 1$ content. Specifically, the $N=1/2$ potential well is of pure $F = 1$ character and the $N = 3/2$ potential well is of mixed $F = 1$ and $F = 2$ character. The vibrational ladder in the $N = 5/2$ potential well, however, cannot be reached, because it has pure $F = 2$ character. Similarly, with an ensemble of $F = 2$ atoms only $N = 3/2$ and $N = 5/2$ vibrational states can be addressed due to their $F=2$ content, but not vibrational states of $N = 1/2$. Therefore, we can now assign the first ladder [$F = 1$ ensemble, blue vertical lines in Fig.$\:$\ref{fig4}(a)] to the $N = 1/2$ potential well, the second ladder [$F = 2$ ensemble, red vertical lines in Fig.$\:$\ref{fig4}(b)] to the $N = 5/2$ potential well, and the third ladder to the $N = 3/2$ potential well. In Fig.$\:$\ref{fig5} we show each measured vibrational ladder for its respective $N$ state together with calculated molecular level energies (see Table \ref{table1} of the Appendix). The potential wells are the same as in Fig.$\:$\ref{fig2}(b), apart from a shift of $1.94\:\text{GHz}\times h$ towards higher energies. The agreement between the measured and calculated vibrational ladders is quite good. The alternation of signal strength is observed for each vibrational ladder and can be explained by the Franck-Condon overlaps, which are in general larger for vibrational wave functions with even symmetry as compared to those with odd symmetry \cite{DeSalvo2015}.   
 
\begin{figure}[t]
	\includegraphics[width=\columnwidth]{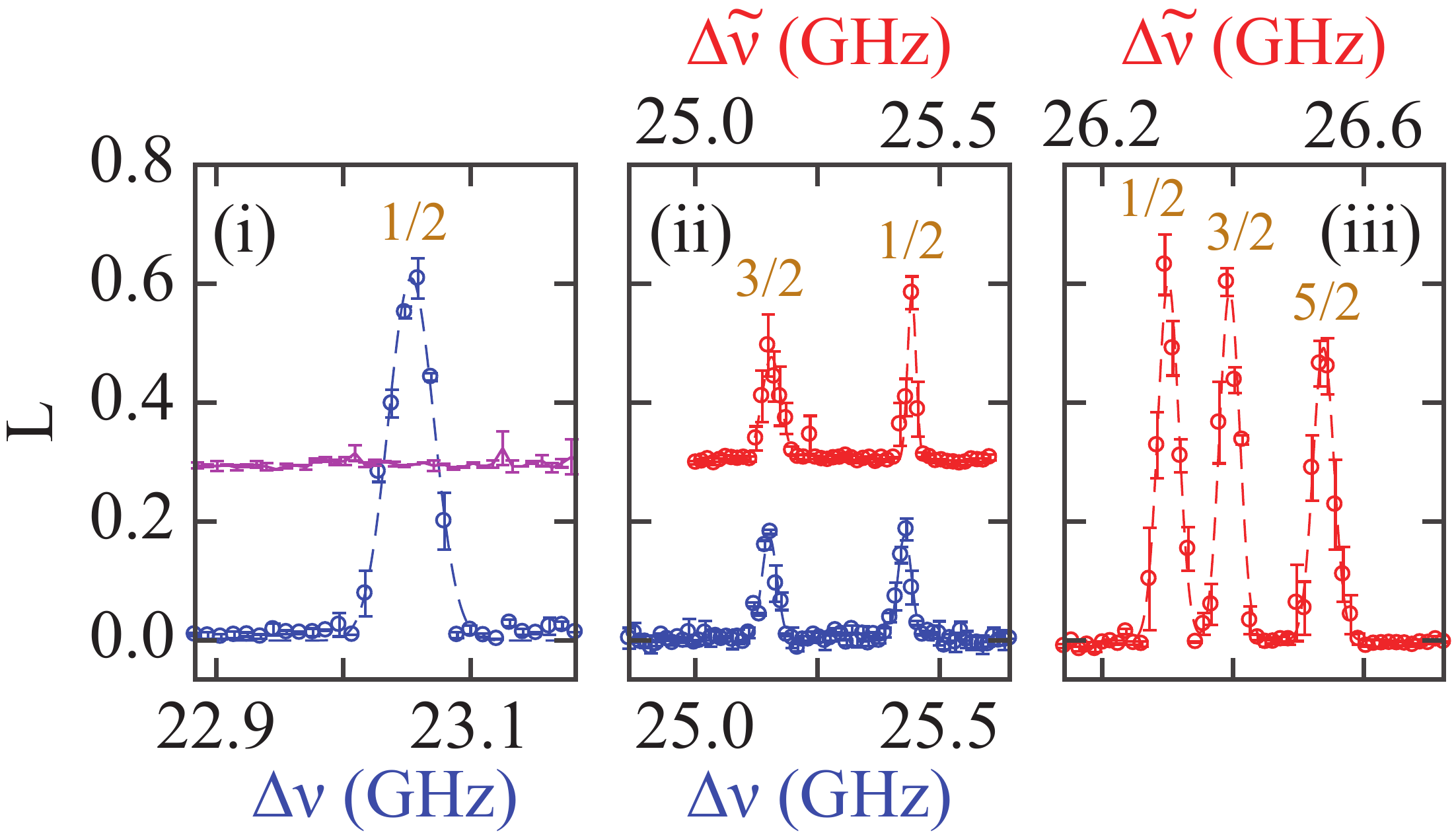}
	\caption{Line multiplets observed for atomic samples initially prepared in the hyperfine state $F=1$ [blue data points for (i) and (ii)] and $F=2$ [red data points for (ii) and (iii)], respectively. The individual lines are labeled with the $|\Omega|$ quantum number of the corresponding assigned molecular state (see also Fig.$\:$\ref{fig5}). In the left panel for the magenta data points the Paul trap was off during the spectroscopy pulse, and no loss signal is visible. Here, the pulse duration of the spectroscopy light was $200\:\text{ms}$ for the $F=1$ data and $300\:\text{ms}$ for the $F=2$ data. For better visibility the magenta data points and also the red data points in the center panel are shifted in the vertical direction by 0.3. The error bars represent the statistical uncertainty. Dashed blue and red lines are the results of Gaussian fits.
	}
	\label{fig6}
\end{figure}

Figure \ref{fig5} reveals that both in the experimental data as well as in the calculations, the vibrational ladders for $N = 3/2$ and $N = 5/2$ exhibit a substructure. For $N = 3/2 $ each vibrational level consists of a doublet and for $N = 5/2$ each vibrational level consists of a triplet. In Fig.$\:$\ref{fig6} we show measurements of these multiplets, which are obtained from high resolution scans. Here, also a singlet line for $N = 1/2$ is presented. The frequency position of each line multiplet [i.e. (i), (ii), and (iii)] is also indicated in the two overview spectra of Fig.$\:$\ref{fig4}. The lines in Fig.$\:$\ref{fig6} are approximately Gaussian shaped and have typical linewidths (FWHM) of a few tens of MHz (see also Sec.$\:$\ref{molecular lines} of the Appendix).

The splitting of the vibrational levels into the multiplets is mainly due to the spin-orbit interaction  $\vec{L}_\text{p} \cdot \vec{S}$. More precisely, each $N=5/2$ vibrational level splits up into three spin components $|\Omega|=\{1/2,\, 3/2,\, 5/2\}$, and each $N=3/2$ vibrational level into a doublet corresponding to $|\Omega|=\{1/2,\, 3/2\}$. Since the $N=1/2$ levels only have $|\Omega|= 1/2 $ they do not split up. These multiplicities agree precisely with our experimental observations and confirm our assignment of the lines.

We now investigate the multiplet splittings in more detail. For the triplets of $N = 5/2$, the observed separation between adjacent lines is typically on the order of about $100$ to $200\:\text{MHz}$. Overall, this is in good agreement with the predictions (see Table \ref{table1} of the Appendix), however there is a systematic increase of the discrepancy for increasing vibrational excitation. Furthermore, the ratio of the energy splittings between the $|\Omega|=1/2$ and $3/2$ components, and the $|\Omega|=3/2$ and $5/2$ components is about $2:3$ on average for the vibrational states, for both experiments and theory. Regarding the line doublets of $N = 3/2$ we typically find splittings of a few hundred MHz. Also here, the discrepancy between measured and calculated splittings increases with vibrational excitation, up to about a factor of two. In addition, we find that the spectroscopy signals for the $N = 3/2$ ladder are in general weaker when working with an $F=2$ atomic ensemble as compared to an $F=1$ ensemble (except for the structure at $\Delta \nu\sim 18.7\:\text{GHz}$). In fact, some of the line doublets could only be detected for $F=1$.

\section{Theory}
\label{sec:calculations}

\subsection{Potential energy curves}
\label{subsec:model}

We determine the molecular PECs by using the electronic Hamiltonian
\begin{equation}
H=H_\text{Ryd} + H_{g} + V\,.
\label{eqn:hamiltonian}
\end{equation}
$H_\text{Ryd}$ describes the interaction of the Rydberg electron in the potential of the ionic Rb$^+$ core, and has eigenstates $\phi_{nljm_j}(\vec{r})$ with energies $E_{nlj}$. The energies $E_{nlj}$ are taken from spectroscopic measurements \cite{Li2003, Han2006} and are utilized as input to analytically determine the long-range behavior (larger than several Bohr radii $a_0$) of $\phi_{nljm_j}(\vec{r})$ in terms of appropriately phase shifted Coulomb wave functions. Knowledge of the wave functions for smaller distances is not necessary for our purpose. $H_\text{g}= A \, \vec{I} \cdot \vec{s}_2$ represents the Hamiltonian of hyperfine interaction in the ground state atom with eigenstates $\ket{F m_F}$, where $A=3.417\:\text{GHz}\times h/\hbar^2$ \cite{Arimondo1977}. The term $V$ describes the interaction between the Rydberg electron and the ground state atom which is largely determined by the orbital angular momentum $\vec{L}_\text{p}$ of the Rydberg electron in the reference frame of the ground state atom. For $L_\text{p}=0$ there is $s$-wave interaction, while $p$-wave interaction is given for $L_\text{p}=1$. We employ a generalized Fermi pseudopotential \cite{Eiles2017, Hummel2017}
\begin{equation}
V= \sum_{\beta} \frac{(2L_\text{p}+1)^2}{2} a(L_\text{p},S,J,k) \frac{\delta(X)}{X^{2(L_\text{p}+1)}} \ket{\beta} \bra{\beta}
\label{eqn:pseudopotential}
\end{equation}
(using atomic units). However, for convenience, we also show in Sec.$\:$\ref{subsec:alternative} of the Appendix how conventional representations of spin-spin and spin-orbit interactions can be derived from this approach, in general. Here, $X=|\vec{r}-\vec{R}|$ is the absolute distance between the Rydberg electron and the ground state atom (see Fig.$\:$\ref{fig1}). The quantum number $J$ corresponds to the angular momentum $\vec{J}=\vec{L}_\text{p}+\vec{S}$, for which the associated magnetic quantum number is denoted by $M_J$. Furthermore, $\beta$ is a multi-index that defines projectors onto the different scattering channels $\ket{\beta}=\ket{L_\text{p} S J M_{J}}$. The interaction strength in each channel depends on the scattering lengths or volumes $a(L_\text{p},S,J,k)=-k^{-(2L_\text{p}+1)}\tan \delta(L_\text{p},S,J,k)$, where $\delta(L_\text{p},S,J,k)$ are phase shifts of an electron with wave number $k$ that scatters off a $^{87}$Rb ground state atom. As a basis for our simulations we employ phase shift data from \cite{Engel2019}. The wave number is calculated via the semiclassical relation $k=\sqrt{2/R-1/n_\text{eff}^2}$. To compute the PECs we use the effective principle quantum number $n_\text{eff}=13.3447$. Please note that neither $\vec{F}$ nor $\vec{j}$ are conserved quantities, since $V$ neither commutes with $H_\text{Ryd}$ nor with $H_\text{g}$.

\begin{table*}
	\begin{tabular}{c c c c c c l}
		\hline
		\hline
		\multicolumn{3}{c}{Scattering channel}&  &  & & \\
		\cline{1-3}
		$\:\:\:\:\:$$L_\text{p}$$\:\:\:\:\:$ & $\:\:\:\:\:$$S$$\:\:\:\:\:$ & $\:\:\:\:\:$$J$$\:\:\:\:\:$ &  & Parameter & & \multicolumn{1}{c}{Mapping} \\
		\hline
		0 & 0 & 0 &  & $\lambda_1$ &  & $ a(0,0,0,k) \mapsto \lambda_1 a(0,0,0,k) + (1- \lambda_1) a(0,1,1,k)$\\
		0 & 1 & 1  & & &  & $a(0,1,1,k) \mapsto a(0,1,1,k)$ \\
		1 & 0 & 1  & & $\lambda_1$ &  & $ a(1,0,1,k) \mapsto \lambda_1 a(1,0,1,k) + (1- \lambda_1) a(1,1,J_\text{avg},k)$ \\
		1 & 1 & 0  & & $\lambda_2$  &  & $a(1,1,0,k) \mapsto \lambda_2 a(1,1,0,k) + (1- \lambda_2) a(1,1,J_\text{avg},k)$ \\
		1 & 1 & 1  & & $\lambda_2$ &  & $a(1,1,1,k) \mapsto \lambda_2 a(1,1,1,k) + (1- \lambda_2) a(1,1,J_\text{avg},k)$ \\
		1 & 1 & 2  & & $\lambda_2$ &  & $a(1,1,2,k) \mapsto \lambda_2 a(1,1,2,k) + (1- \lambda_2) a(1,1,J_\text{avg},k)$ \\
		\hline
		\hline
	\end{tabular}
	\caption{Overview of the scattering lengths or volumes $a(L_\text{p},S,J,k)$ that are modified via control parameters $\lambda_1$ and $\lambda_2$ in order to study the splitting mechanisms in Fig.$\:$\ref{fig7}. The scattering channels are given in terms of the quantum numbers $L_\text{p}$ ($L_\text{p}=0$: $s$-wave scattering; $L_\text{p}=1$: $p$-wave scattering), $S$ ($S=0$: singlet scattering; $S=1$: triplet scattering), and $J=\{0,\,1,\,2\}$. \label{tab:map}}
\end{table*}

In general, for our calculations the Hilbert space is restricted to a subset of Rydberg states in the spectral region of interest, as described in Sec.$\:$\ref{sec:HilbertSpace} of the Appendix. The PECs obtained by taking into account both $s$-wave and $p$-wave interactions in Eq.$\:$(\ref{eqn:hamiltonian}) are shown in Fig.$\:$\ref{fig2}. The relevant curves are characterized in Fig.$\:$\ref{fig2}(b) by $N$ and $|\Omega|$. $N$ provides the correct multiplicity, however, strictly speaking, $N$ is not a good quantum number. Instead, $\Omega$, represents a good quantum number, and it is appropriate to further discriminate the PECs. Please note that $\Omega$ is not the projection quantum number of $N$.

\begin{figure}[b]
	\includegraphics[width=\columnwidth]{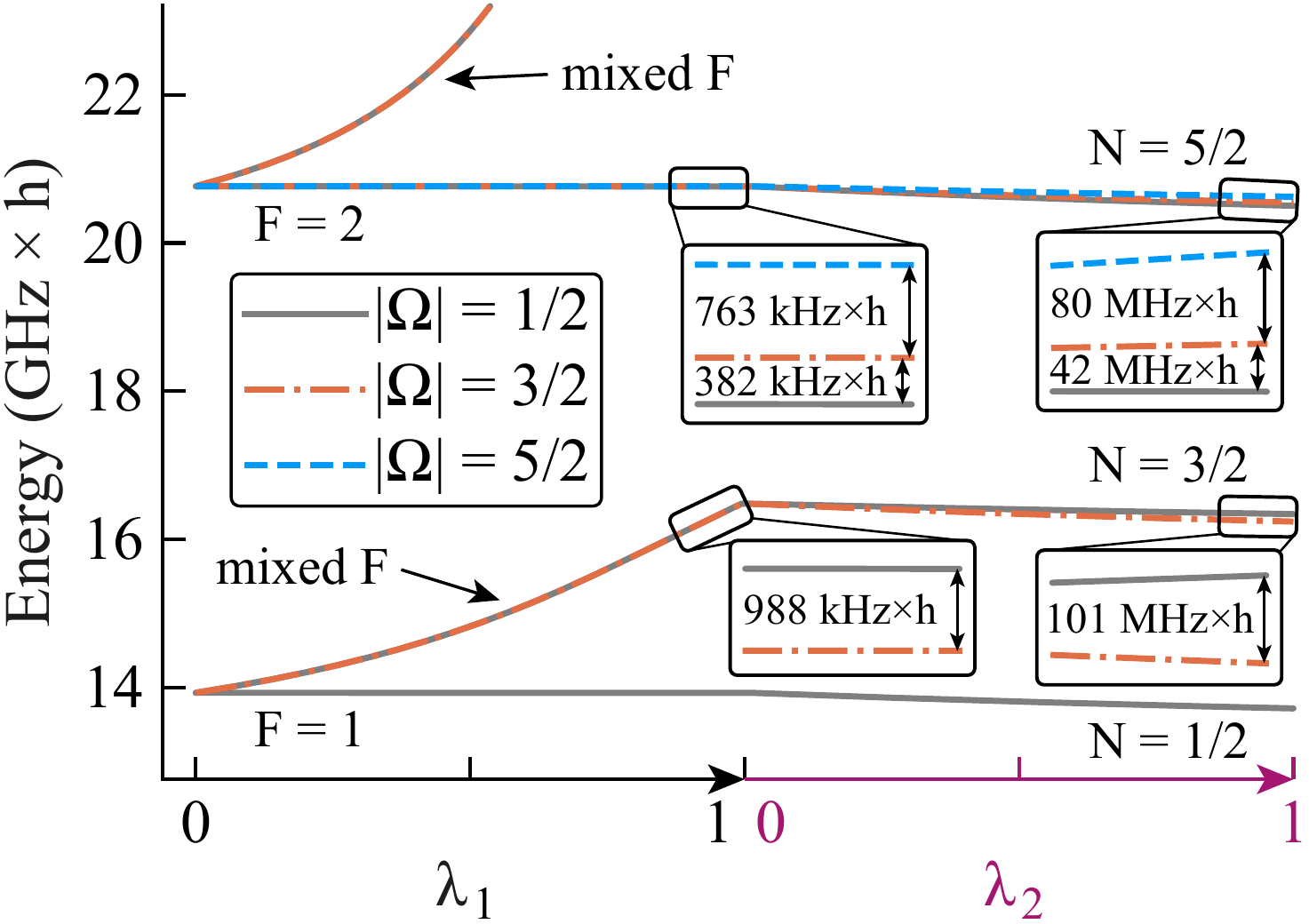}
	\caption{Values of the molecular PECs at an internuclear distance $R=265\:a_0$ as a function of the interaction control parameters $\lambda_1$ and $\lambda_2$ (see also Table \ref{tab:map}). On the left, where $\lambda_1=\lambda_2=0$, interaction is identical for singlet and triplet states. $s$-wave and $p$-wave interactions are, however, not identical. When going to the right $\lambda_1$ and $\lambda_2$ are subsequently turned on. Parameter $\lambda_1$ modifies the singlet $s$-wave and $p$-wave scattering and introduces a splitting of the lines in two respects. First, branches of mixed $F$ character separate from branches of pure $F$ character. Second, the $|\Omega|$ components within these branches slightly split off from each other. Parameter $\lambda_2$ introduces a $\vec{L}_\text{p}\cdot\vec{S}$ type of interaction. This enhances the $|\Omega|$ splittings by about two orders of magnitude.}
	\label{fig7}
\end{figure}

\subsection{Comparison of spin-spin and spin-orbit interactions}
\label{subsec:LSmodel}

In the following, we investigate in detail the reasons for the splitting of the PECs with a given $N$ quantum number into the various $|\Omega|$ components. It will turn out that the $\vec{L}_\text{p}\cdot \vec{S}$ interaction is by far the dominant mechanism. For our investigation, we introduce two control parameters $\lambda_1$ and $\lambda_2$. These allow for relative tuning of different scattering channels by modifying the scattering lengths or volumes $a(L_\text{p},S,J,k)$, which helps us to gain insight about the role of relevant interactions. The mapping is summarized in Table \ref{tab:map}. We analyze the impact of the individual control parameters on the PECs for an internuclear distance of $265\:a_0$, which roughly corresponds to the locations of the minima of the potential wells in Fig.$\:$\ref{fig2}(b). This choice is motivated by the positions of the barycenters of the vibrational wave functions. The results are shown in Fig.$\:$\ref{fig7}.

When $\lambda_1=\lambda_2=0$, the electron-atom interaction $V$ is insensitive to the total electronic spin $\vec{S}$ and the interaction can be simplified to \cite{Greene2000,Hamilton2002,note}
\begin{equation}
V= 2 \pi a_s(k) \delta(\vec{R}-\vec{r}) + 6 \pi a_p(k) \overleftarrow{\nabla}_{\vec{r}} \cdot \delta(\vec{R}-\vec{r}) \overrightarrow{\nabla}_{\vec{r}}
\end{equation}
with $a_s(k)=a(0,1,1,k)$, $a_p(k)=a(1,1,J_\text{avg},k)$, and $\vec{R}=R \hat{e}_z$. The phase shift for $J_\text{avg}$ corresponds to the situation, in which the $\vec{L}_\text{p}\cdot\vec{S}$ coupling is neglected. For the resonance energy $E_r^{\text{avg}}$ associated with this phase shift we use the value $E_r^{\text{avg}}=26.6\:\text{meV}$ taken from \cite{Engel2019}. Figure \ref{fig7} shows that for this case there is no splitting of the PECs for both the $F=1$ and the $F=2$ branch.

We now let $\lambda_1>0$, while keeping $\lambda_2=0$. The parameter $\lambda_1$ introduces a difference in the singlet and triplet scattering lengths or volumes. As a consequence, typically, a splitting of each of the $F = 1,\,2$ branches occurs, i.e. a separation of states with mixed $F$ character from those with pure $F=1,\,2$ character is obtained. The energy differences between pure and mixed $F$ states change as a function of the internuclear distance (see Fig.$\:$\ref{fig10} of the Appendix). For the specific choice of $R=265\:a_0$ these are on the order of several $\text{GHz}\times h$ for $\lambda_1=1$ in Fig.$\:$\ref{fig7}. Additionally, the parameter $\lambda_1$ lifts the energetic degeneracy of the different $|\Omega|$ components for the individual $F$ branches. However, the introduced splittings of the $|\Omega|$ states are below $1\:\text{MHz}\times h$ for $R=265\:a_0$, and therefore very small. Further information on the separation of $F$ branches and the impact of $\lambda_1$ on $|\Omega|$ components is given in Sec.$\:$\ref{splitting mechanisms} of the Appendix.

For the regime $\lambda_1=1,\,\lambda_2>0$ the full interaction introduced in Eq.$\:$(\ref{eqn:pseudopotential}) is realized by including the $J$ dependency of the $p$-wave triplet scattering. The physical origin of the $J$ dependency is $\vec{L}_\text{p} \cdot \vec{S}$ spin-orbit coupling. Each $J$ channel ($J=\{0,\,1,\,2\}$) is associated with a characteristic energy $E_r^J$ where the $p$-wave shape resonance occurs. For our scattering phase shifts these values are $E_r^{J=(0,\,1,\,2)}=(24.4,\,25.5,\,27.7)\:\text{meV}$, respectively \cite{Engel2019}. We note that the energies $E_r^J$ follow the Land\'{e} interval rule. Thus, electronic triplet states of different $J$ experience different interaction strength for any given internuclear separation. Figure \ref{fig7} shows that this leads to additional, strikingly large splittings of the $|\Omega|$ components, on the order of tens of $\text{MHz}\times h$ for the $F=2$ branch and up to about $100\:\text{MHz}\times h$ for the lower mixed $F$ branch. This is by about two orders of magnitude larger than the splitting due to $\lambda_1$ spin-spin interaction. Therefore, we conclude that the shapes of the observed multiplet substructures are almost entirely determined by $\vec{L}_\text{p} \cdot \vec{S}$ spin-orbit interaction.

We note, that in general, the $|\Omega|$ splittings depend on the internuclear distance $R$ due to the energy dependence of the scattering lengths or volumes $a(L_\text{p},S,J,k)$ as well as the spatial variation of the Rydberg electron wave function. This can be seen, e.g., in the PECs of Fig.$\:$\ref{fig2}(b). For example, when considering the $N=3/2$ doublet, within the potential wells, for smaller values of $R$ (i.e. closer to the $p$-wave shape resonance) the $|\Omega|=1/2$ and $|\Omega|=3/2$ states are further energetically separated from each other than for higher values of $R$. This is due to the fact that resonant $p$-wave interactions amplify the effect of $\vec{L}_\text{p} \cdot \vec{S}$ coupling. In order to check for consistency we have varied the internuclear separation around the value of $R=265\:a_0$ used for Fig.$\:$\ref{fig7}. A corresponding analysis reveals that the ratios of $|\Omega|$ splittings introduced by parameters $\lambda_1$ and $\lambda_2$ are robust, i.e. over the whole potential wells of Fig.$\:$\ref{fig2}(b) $\vec{L}_\text{p} \cdot \vec{S}$ coupling still remains the dominant interaction that energetically separates the $|\Omega|$ components. Only when going to the left of the barriers very close to the $p$-wave shape resonance (e.g. at an internuclear distance of about $220\:a_0$) does the relative impact of the parameter $\lambda_1$ increase significantly.

\section{Conclusions and Outlook}
\label{sec:conclusion}

In conclusion, we find evidence for spin-orbit dependent scattering of an electron from a neutral atom. The scattering takes place within an ultralong-range Rydberg molecule which represents a micro laboratory for low-energy scattering experiments. We observe the spin-orbit interaction directly and quantitatively in terms of bound state level splittings of the ultralong-range Rydberg molecule. These level splittings are particularly large in the chosen parameter regime close to a $p$-wave shape resonance which enhances the effect of spin-orbit coupling on the molecular structure. Model calculations agree well with our experimental data and allow for assigning all relevant spin states to observed levels.

Having obtained a good understanding of the complex spin-couplings and level-structures of the ultralong-range Rydberg molecules, it is now possible to study interesting spin and wave packet dynamics in these systems. In fact, for Rb$_2$ molecules having principal quantum numbers in the vicinity of $n=16$, our calculations predict that the level crossings of the butterfly state with the $P$ state curves will give rise to non-trivially coupled potential energy landscapes where, e.g., non-adiabaticity effects (such as the breakdown of the Born-Oppenheimer approximation) and interesting tunneling effects, can be studied. The ultralong-range Rydberg molecule will then become an even more versatile micro laboratory for fundamental quantum dynamics aspects \cite{Koeppel1984, Worth2004}.
 
In addition, the presented observation and interpretation of spin structures sets a basis for further high precision Rydberg spectroscopic studies. These will allow for testing the limits of the theoretical understanding and modeling of the Rydberg system, in general. In fact, it might turn out that the effective pseudopotential approach is not adequate to fully describe all relevant interactions, as it suffers from limited accuracy due to convergence issues \cite{Eiles2017,Fey2015}. Precision spectroscopy data will therefore spark increased efforts, e.g. in the development of appropriate $R$-matrix methods \cite{Tarana2016} or the inclusion of spin interactions in Green's function approaches, to obtain a consistent theoretical treatment. 
   
Finally, our results on spin-spin and spin-orbit coupling are helpful for current research activities regarding polyatomic many-body systems (see, e.g., \cite{Schlagmueller2016b, Fey2016, Eiles2016, Schmidt2016, Ashida2019}) due to the fundamental importance of pairwise interactions between two atoms.

\section*{Acknowledgments}
We thank Herwig Ott for valuable discussions. This work was supported by the German Research Foundation (DFG) within the priority program "Giant Interactions in Rydberg Systems" [DFG SPP 1929 GiRyd (projects HE 6195/3-1, PF 381/17-1, and SCHM 885/30-1)], and under DFG project PF 381/13-1. M.D. acknowledges support from Universit\"{a}t Ulm and Ulmer Universit\"{a}tsgesellschaft (UUG) through a Forschungsbonus grant. F.M.~acknowledges support from the Carl Zeiss Foundation and is indebted to the Baden-W\"{u}rttemberg Stiftung for financial support by the Eliteprogramm for postdocs.

\section*{Appendix}
\makeatletter
\def\p@subsection{}
\makeatother

\subsection{Electronic spin character of the potential energy curves}
\label{sec:spincharcter}

\begin{figure}[b]
	\includegraphics[width=\columnwidth]{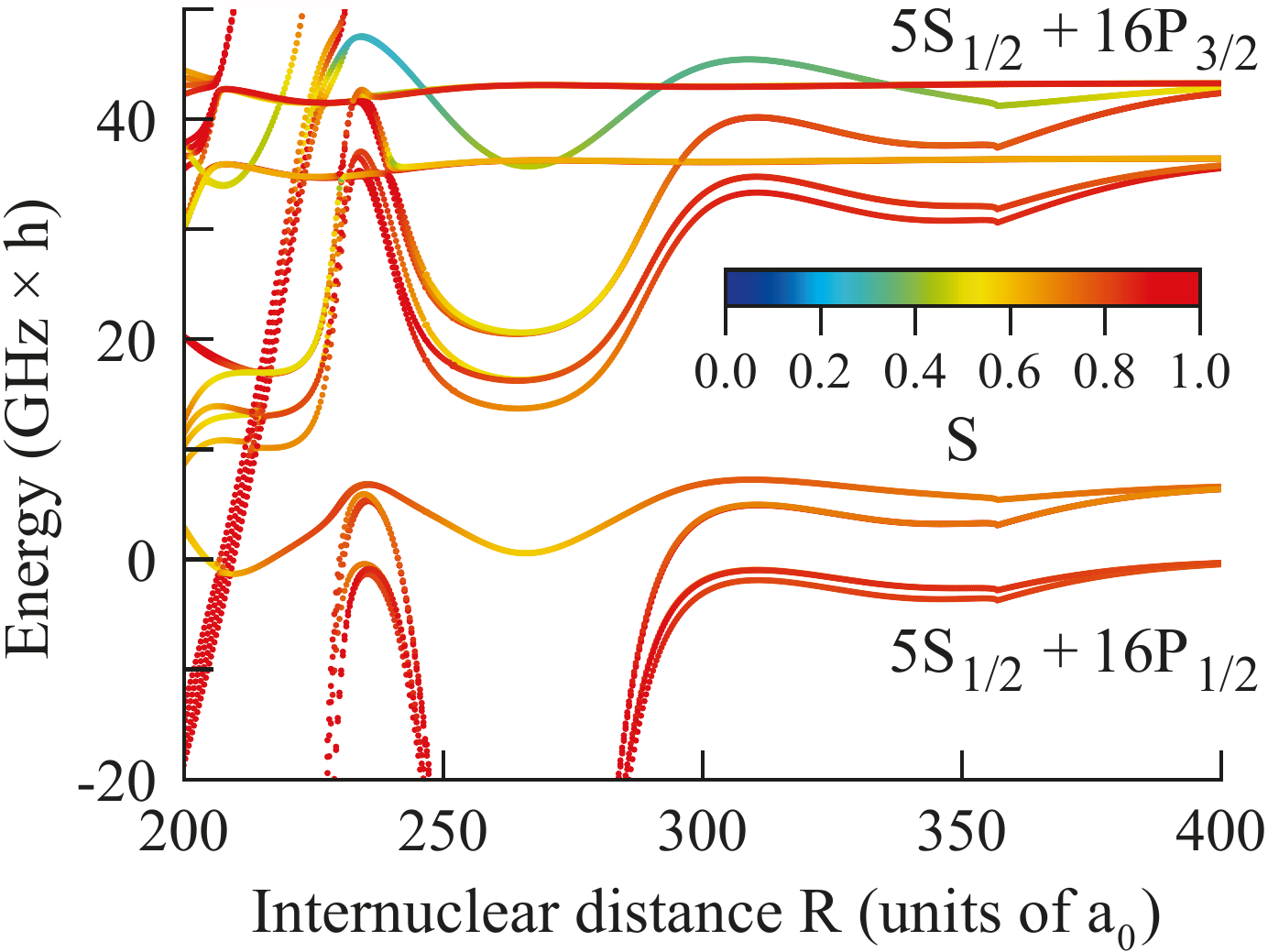}
	\caption{The molecular PECs correlated to the $5S_{1/2}+16P_j$ atomic asymptotes for $j\in\{1/2,\,3/2\}$ and different hyperfine states $F\in \{1,\,2\}$ of the $5S_{1/2}$ atom. Here, the color code represents the expectation value of the quantum number $S$ of the total electronic spin.
	}
	\label{fig8}
\end{figure}

Figure \ref{fig8} shows the same PECs as Fig.$\:$\ref{fig2} of the main text, however, the color coding gives the expectation value of the total spin $S$. Interestingly, the PECs differ quite substantially in their $S$ spin character despite the fact that the hyperfine character $F$ is nearly constant for a given set of curves that belong to the same $N$.

We note that the cusps in the outer wells of the PECs around $R=360\:a_0$, which are visible e.g. in Fig.$\:$\ref{fig8} and Fig.$\:$\ref{fig10}, occur due to the non-analytic behavior of the wave number $k$ close to the classical turning point, where $k$ becomes zero.

\subsection{Photoassociation setup}
\label{PA setup}
The photoassociation laser operates at wavelengths of around $302\:\text{nm}$. The laser light is generated by a frequency-doubled cw dye laser with a narrow short-time linewidth of a few hundred kilohertz. The laser is frequency-stabilized to a wavelength meter (High Finesse WS7) which is repeatedly calibrated to an atomic $^{87}$Rb reference signal at a wavelength of 780$\:$nm in intervals of hours. We achieve a shot-to-shot frequency stability of below $\pm10\:\text{MHz}$ for the $302\:\text{nm}$ light.

A multi mode optical fiber is used to transfer the UV light to the experimental table. At the location of the atoms the spectroscopy beam has a waist ($1/e^2$ radius) of about $1.5\:\text{mm}$ and the power is typically in the range of $4$ to $10\:\text{mW}$. The light pulse has a rectangular shape and the atoms are exposed to the laser radiation for a duration on the order of $0.1$ to $1\:\text{s}$.

\subsection{Atomic lines}
\label{atomic lines}

The strong resonance lines marked with horizontal black arrows in the spectra (a) and (b) of Fig.$\:$\ref{fig4} correspond to the atomic transitions towards $16P_{1/2}$ and $16P_{3/2}$. Here, the atom loss of the atomic cloud is close to 100$\%$. The $16P_{1/2}$ line is located at $\Delta \nu=0$ in (a) and at $\Delta \tilde{\nu} \approx \nu_\text{hfs} = 6.835\:\text{GHz}$ in (b) which corresponds to the ground state hyperfine splitting. For the excited Rydberg $P$ state the hyperfine splitting can be neglected. The asymmetric tail on the red side of each atomic resonance line arises from the Stark effect due to the electric fields of both the Paul trap and the trapped ions (see also \cite{Haze2019,Ewald2018,Engel2018}). The strong resonance lines marked with vertical black arrows in the spectra (a) and (b) of Fig.$\:$\ref{fig4} also correspond to transitions towards the atomic $16P_{1/2}$ and $16P_{3/2}$ states. These lines are shifted by about $\pm\nu_\text{hfs}$ relative to the atomic resonance lines marked with horizontal black arrows. Apparently, each of the prepared $F = 1$ ($F=2$) samples is not 100\% pure but contains a fraction of atoms in the other spin state $F = 2$ ($F=1$), respectively. Although these admixed fractions are possibly on the percent level or less they still can give rise to large signals due to the non-linear behavior of the atomic loss, as discussed in the main text.

\subsection{Mining of experimental data}
\label{Collection}

\begin{figure}[t]
	\includegraphics[width=\columnwidth]{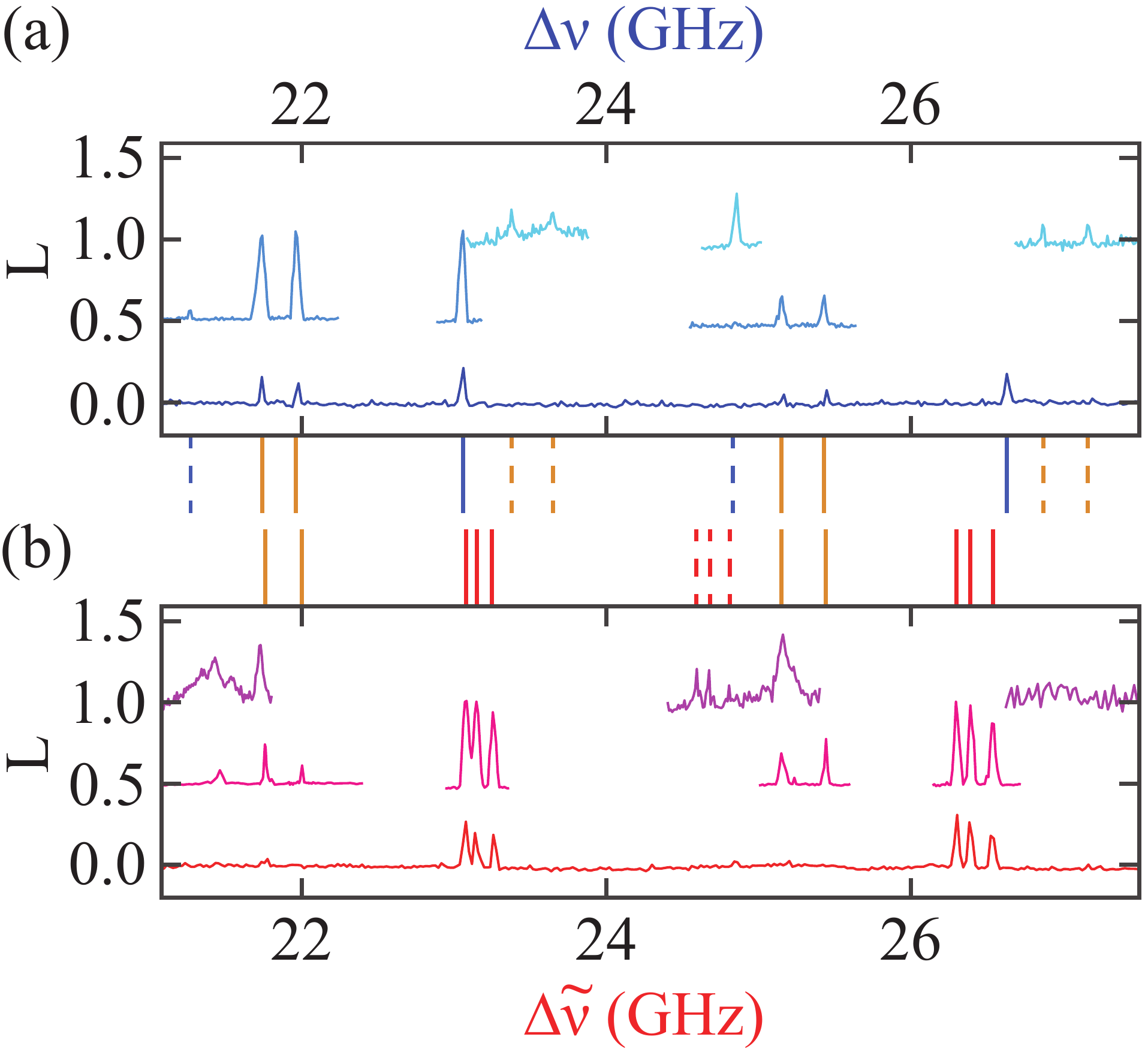}
	\caption{Measured series of line multiplets for atomic samples initially prepared in the hyperfine state $F=1$ (a) and $F=2$ (b). The blue and red data curves in (a) and (b), respectively, are the same as in Fig.$\:$\ref{fig4}. All other spectra are obtained for individually optimized experimental parameters to locally increase the signal-to-noise ratio. For better visibility, light blue (cyan) data in (a) are shifted in the vertical direction by 0.5 (1.0), as well as the magenta (purple) data in (b). Blue and orange (red and orange) vertical lines at the bottom of plot (a) [at the top of plot (b)] indicate the frequency positions of measured resonances belonging to line singlets and line doublets (line triplets and line doublets) in (a) and (b), respectively. Orange color corresponds to light gray in grayscale versions. The alternating signal strength behavior for each multiplet series is illustrated by the line style, where solid (dashed) lines represent strong (weak) signals.}
	\label{fig9}
\end{figure}

In general, we have various parameters available to tune signal strengths for the detection of ultralong-range Rydberg molecules. These are the intensity and pulse duration of the spectroscopy light, but also the ionic micromotion energy and the time for interaction between trapped ions and neutral atoms. Figure \ref{fig9} shows qualitatively how signals change when we vary these parameters, as indicated by different line colors and scan ranges. The dark blue data in (a) and the red data in (b) are zooms into Figs.$\:$\ref{fig4}(a) and (b), respectively. Additional resonance lines which are not visible in these two spectra, can be revealed after individual parameter optimization. The blue, orange, and red vertical lines represent the center frequency positions of the measured resonances of line singlets, line doublets, and line triplets. Solid and dashed vertical lines mark strong and weak signals, respectively. They alternate between adjacent vibrational states for each of the three observed ladders. Within the given frequency range we observe almost the complete series of expected resonances for each multiplet structure. Only the weak line doublets for $F=2$ are missing in Fig.$\:$\ref{fig9}(b) (see purple data scan at around $\Delta \tilde{\nu}=27\:\text{GHz}$). An overview of all observed (and calculated) molecular level positions is provided in Table \ref{table1}.

\subsection{Linewidths of molecular signals}
\label{molecular lines}

The measured linewidths of several tens of MHz are more than one order of magnitude larger than expected from the natural lifetimes of the molecular states. These natural lifetimes should be on the order of that of the atomic $16P_{3/2}$ Rydberg state, for which a value of about $4\:\upmu\text{s}$ is predicted \cite{Theodosiou1984}. The observed large linewidths of the molecular lines might be explained by the uncertainty of the UV photoassociation laser of about $\pm10\:\text{MHz}$ and due to the Stark effect. In our experimental scheme an ion trap is used and therefore dc and ac, position-dependent electric fields are present. A detailed analysis of electric dipole moments and a simulation of the impact of the Stark effect on linewidths of molecular signals needs to be done in future work. Finally, we note that also limitations in the lifetime arising from the ionization of molecules subsequent to their formation can play a role.

\subsection{Restricting the Hilbert space for numerical calculations}
\label{sec:HilbertSpace}
The Hamiltonian $H$ is constructed in a finite basis set that includes the 15$S$, 16$S$, 17$S$, 14$P$, 15$P$, 16$P$, 13$D$, 14$D$, and 15$D$ states, and the hydrogenic states with higher orbital angular momenta $l\geq3$ with principle quantum numbers $n=12$, $n=13$, and $n=14$. All these states are considered with all possible total angular momenta $j$, while the projections $m_j$ are truncated to include $|m_j|\leq 3/2$. According to the choice of the molecular axis lying on the $z$ axis, states with $|m_j|>3/2$ do not interact with the ground state atom. Additionally, the nuclear and electronic spins of the ground state atom are taken into account completely ($m_I=\{\pm1/2, \pm3/2\}$ and $m_{s_2}=\pm1/2$). Note, that placing the perturber onto the $z$ axis significantly reduces the basis set. Since the scattering interaction $V$ vastly exceeds the Zeeman energy for any magnetic fields occurring due to the experimental setup, the atomic orbitals align along the internuclear axis. This is different, however, when the interaction with an external field is comparable to or larger than the scattering interaction \cite{Krupp2014}. Alternative approaches to derive the PECs that circumvent a finite basis set are Green's function methods employed for example in \cite{Chibisov2002, Khuskivadze2002}. However, these approaches do not incorporate spin interactions that are crucial for the interpretation of our results. Nevertheless, we used a Green's function approach and a reduced spin model which neglects fine and hyperfine structure to find the optimal basis size. The corresponding basis was then employed for the full model calculations.

\subsection{Discussion of splitting mechanisms}
\label{splitting mechanisms}

\begin{figure}
	\includegraphics[width=\columnwidth]{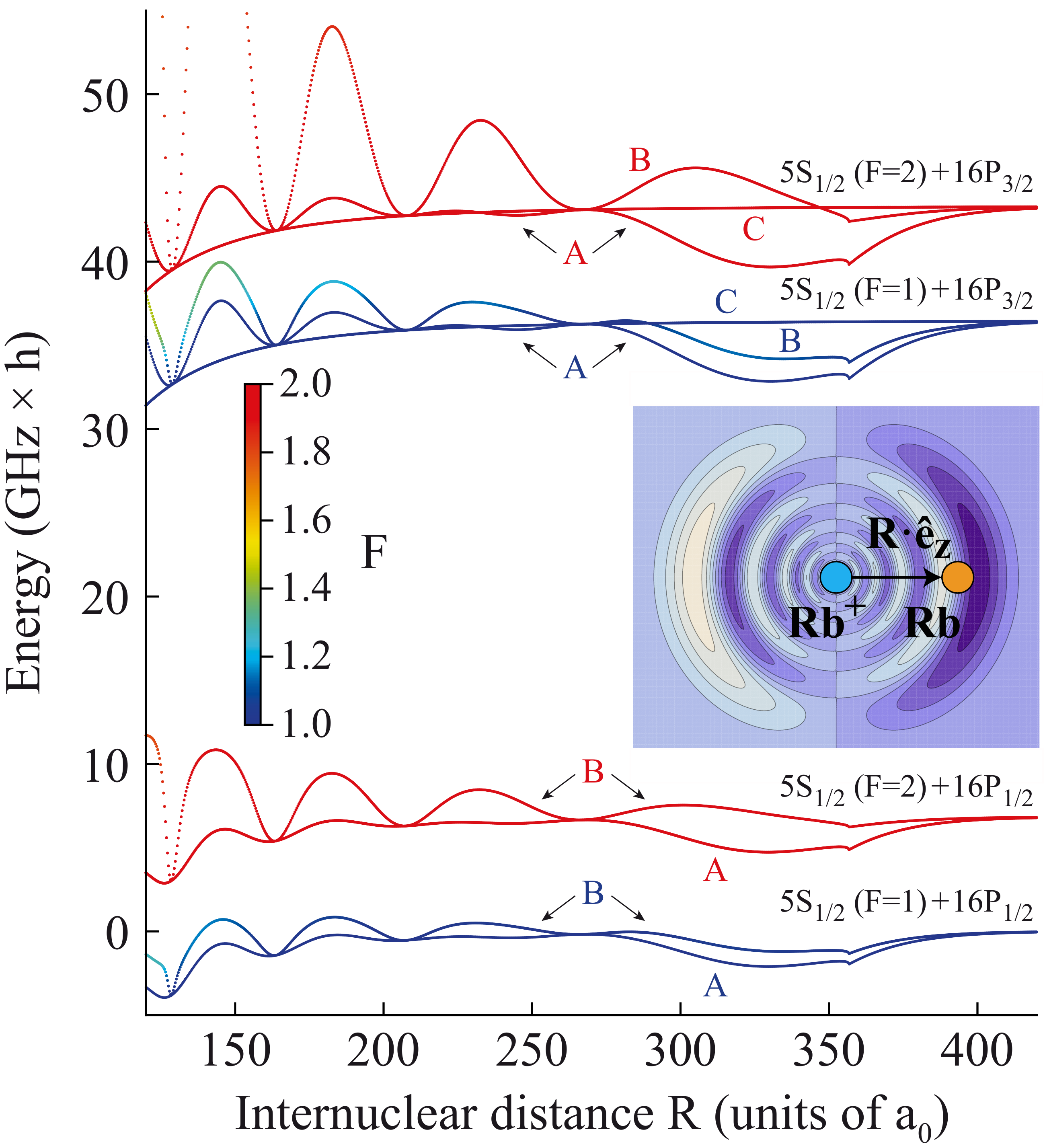}
	\caption{The molecular Born-Oppenheimer potentials when $s$-wave interactions between the electron and the ground state atom are taken into account but $p$-wave interactions are neglected. The colors of the curves indicate the expectation value of the $F$ quantum number of the ground state atom. From each atomic asymptote two oscillatory potentials emerge. They have  $|m_j|=1/2$ and feature the nodes of the electronic wave function. The deeper PECs (marked with A) are associated with pure triplet scattering and the shallower PECs (marked with B) are associated with mixed singlet/triplet scattering. The PECs labeled with C have $|m_j|>1/2$ and do not show an oscillatory behavior. In the inset a sketch of an ultralong-range Rydberg molecule is shown. It consists of a Rb$^+$ ionic core, an electronic Rydberg $16P$ state orbital, and a ground state Rb atom which is located at position $\vec{R}=R\hat{e}_z$ relative to the ionic core. The $16P$ electronic orbital is given in a contour plot representation.
	}
	\label{fig10}
\end{figure}

In order to recall the molecular setup, the inset of Fig.$\:$\ref{fig10} shows the electronic $16P$ orbital of the Rydberg atom, which overlaps with the ground state atom at distance $R$. To a first approximation the interaction between the ground state atom and the Rydberg electron can be modeled by a short-range, $s$-wave Fermi-type pseudopotential. In Fig.$\:$\ref{fig10} the Born-Oppenheimer PECs are shown, when the $p$-wave interaction is neglected in Eq.$\:$(\ref{eqn:hamiltonian}), i.e. $a(L_\text{p} = 1,S,J,k) = 0$. Using this simplified situation aids convenient discussion in the following.

The oscillatory behavior of the PECs in Fig.$\:$\ref{fig10} reflects the radial wave function of the Rydberg electron. Here, the separation of states with mixed $F$ character from those with pure $F=1,\,2$ character as a function of the internuclear distance $R$ can directly be seen for $R\lesssim 370\:a_0$. Each asymptote breaks up into two oscillatory PECs, marked with A and B. There is an additional, non-oscillatory PEC, marked with C, for each $P_{3/2}$ asymptote. These C PECs correspond to Rydberg states with $m_j = \pm 3/2$ which do not undergo $s$-wave interaction, because the ground state atom on the $z$ axis is located at the node of the $|m_l| = 1 $ electronic orbital. The remaining interaction of the C PECs in Fig.$\:$\ref{fig10} is then solely through the attractive $1/R^4$ polarization potential due to the Rb$^+$ ionic core.

In the literature \cite{Anderson2014a, Anderson2014b, Sassmannshausen2015, Boettcher2016} each pair of oscillatory PECs is subclassified into a 'deeper' curve (A) and a 'shallower' curve (B) which are sometimes also labeled 'triplet' and 'mixed', respectively. However, the deeper curves are not pure triplet states due to Rydberg fine structure. We explain how this is possible with the following example. We consider the electronic state of a Rydberg atom in a $P$ state with total orbital angular momentum $j=1/2$ and projection $m_j=1/2$ and a ground state atom in a polarized nuclear spin state $F=2$ and $m_F=2$,
\begin{equation}
\left(
\begin{array}{c}
\psi_{m_l=0,\uparrow} (\vec{r})\\
\psi_{m_l=1,\downarrow} (\vec{r})\\
\end{array}
\right) \otimes \ket{F=2, m_F=2}\,.
\label{eqn:spinor}
\end{equation}
In first order perturbation theory (with respect to weak $s$- and $p$-wave interactions) this state must be an eigenstate of the Hamiltonian, as it is the only possible realization of an $\Omega = 5/2$ state in the Hilbert subspace considered here. The spin-up component has $m_l=0$ and the spin-down component has $m_l=1$. Together with the spin-stretched ground state atom, the spin-up component forms a pure spin triplet. Therefore, this component does not interact in the $s$-wave singlet channel. The spin-down component is a mixed singlet/triplet state. However, it still does not interact in the $s$-wave singlet channel because $m_l=1$ and hence the ground state atom is located at the node of the electronic wave function. Thus, despite the fact that the state of Eq.$\:$(\ref{eqn:spinor}) has a singlet component, it is insensitive to $s$-wave singlet interaction.

The state of Eq.$\:$(\ref{eqn:spinor}) is just one example for the many degenerate eigenstates associated with the deep PECs. The degeneracy of the deep and shallow PECs can be obtained with the help of the spin operator $\vec{N}^2$, where $\vec{N}=\vec{S}+\vec{I}= \vec{s}_1+\vec{F}$. We note that $\vec{N}^2$ does not commute with $H_\text{Ryd}$ due to the Rydberg fine structure, however, it is still useful for labeling the scattering channels, as we show in the following. The basis states of a given $F$ branch all have a form similar to that of the state of Eq.$\:$(\ref{eqn:spinor}). Within the vector space spanned by these basis states, we want to determine the dimension of the subspace that is susceptible to singlet $s$-wave interaction. Since the $|m_l| = 1$ component of a basis state does not contribute to $s$-wave interaction, we only consider its $m_l = 0$ component, of which the spin can be up or down. Thus, the problem can be reduced to determining the dimension of the formed $S = 1$ subspace when coupling an electronic spin $\vec{s_1}$ to the angular momentum  $F$ manifold  where $\vec{F} = \vec{I} + \vec{s_2} $. For this, we divide up the resulting new manifold into subspaces with good quantum number $N$. Since $\vec{N}^2$ commutes with both $\vec{S}^2$ and $\vec{F}^2$, this will help us in sorting out the spin structure. We note, however, that $\vec{S}^2$ and $\vec{F}^2$ do not commute. For $F = 2$, $N$ can be $N = 5/2$ or $N =  3/2 $. Since the $N = 5/2$ subspace must have $S = 1$ it belongs to branch A. With the help of Wigner $6j$ coefficients one can show that the $N = 3/2$ subspace, however, contains states with singlet and triplet character and therefore belongs to branch B. Similarly, for $F = 1$, we have the subspaces
$N = 3/2$ and $N =  1/2 $. Here, $N = 1/2$ goes along with $S = 1$ and thus belongs to branch A, whereas $N = 3/2$ includes both $S$ characters and belongs to branch B. As the difference in the singlet and triplet $s$-wave interactions becomes larger, the two $N = 3/2$ manifolds of the $F = 1$ and $F = 2$ branches start mixing. The degree of $F$ mixing depends on the relative strength of the differential singlet/triplet $s$-wave interaction and the hyperfine interaction $H_\text{g}$. The $F$ mixing due to the presence of a singlet $s$-wave scattering channel is essential for the spin flip effect observed in \cite{Niederpruem2016b}. $N$ reproduces the multiplicities for the PECs, which are visible in Fig.$\:$\ref{fig2}(b) and Fig.$\:$\ref{fig7}. For the $F=2$ asymptote, the deep curve corresponds to $N=5/2$ and has six degenerate states of pure $F=2$ character. For the $F=1$ asymptote, the deep curve corresponds to $N=1/2$ and has two degenerate states of pure $F=1$ character. The shallow curves of both, the $F=1$ and the $F=2$ asymptotes correspond to $N=3/2$ and have four degenerate states of mixed $F=1$ and $F=2$ character each. While this regime of interactions is sufficient to describe the PECs at the outer potential wells (in this case for $R>300\:a_0$), additional $p$-wave related interactions become important for smaller internuclear separations, which are also relevant in the parameter regime of $\lambda_1$.

The difference between the singlet and triplet channels introduced via parameter $\lambda_1$ affects in particular the spinor components with $m_l=\pm1$, which only probe the $p$-wave interaction but not the $s$-wave interaction. Although the state with $\Omega=5/2$ of Eq.$\:$(\ref{eqn:spinor}) is of pure triplet character in its $m_l=0$ component, it is of mixed singlet/triplet character in its $m_l=1$ component, as discussed before. As a consequence, it will experience a first order level shift. States with a different $|\Omega|$ have different mixing ratios and will exhibit different shifts. Therefore, the spin-selective $p$-wave interaction generally leads to a splitting of $|\Omega|$ states. This splitting, however, arises only due to the Rydberg fine structure and is, hence, not visible in $S$ state ultralong-range Rydberg molecules recently studied \cite{Hummel2018}, since they do not exhibit such a kind of fine structure related to $|m_l|=1$.

\subsection{Alternative representation of spin-spin and spin-orbit interaction}
\label{subsec:alternative}

The pseudopotential that models the interaction between the Rydberg electron and the ground state atom is given in Eq.$\:$(\ref{eqn:pseudopotential}) of the main text. Our aim is now to rewrite the given interaction potential in terms of operators such as $\vec{s}_1 \cdot \vec{s}_2$ and $\vec{L}_\text{p} \cdot \vec{S}$, respectively.

First, we consider the example of pure $s$-wave scattering, i.e. $L_\text{p}=0$ and therefore $J=S$ and $M_J=M_S$. The corresponding expression of Eq.$\:$(\ref{eqn:pseudopotential}) is compared to the ansatz
\begin{equation}
V_{L_\text{p}=0}= \left[c_1 \hat{1}+ c_2\vec{s}_1 \cdot \vec{s}_2 \right] \frac{\delta(X)}{2X^2} |L_\text{p}=0 \rangle \langle L_\text{p}=0| \,.
\label{eqn:spin spin}
\end{equation}
For this, Eq.$\:$(\ref{eqn:spin spin}) is represented as a $4 \times 4$ matrix in the basis $|S,M_S \rangle$. From the comparison we find that both expressions are identical if $c_1=[a(0,0,0,k)+3a(0,1,1,k)]/4$ and $c_2=a(0,1,1,k)-a(0,0,0,k)$. As expected, there is no spin-spin coupling, i.e. $c_2=0$, when the $s$-wave singlet and triplet scattering lengths or volumes $a(0,0,0,k)$ and $a(0,1,1,k)$ are equal. Furthermore, $c_1$ corresponds to the averaged $s$-wave scattering length or volume.

Now, we turn to a treatment of spin-orbit coupling. Since spin-orbit interaction only takes place in the $p$-wave triplet channel only the subspace with $L_\text{p}=1$ and $S=1$ has to be considered in Eq.$\:$(\ref{eqn:pseudopotential}). The resulting expression is compared to the ansatz
\begin{align}
V_{L_\text{p}=S=1} =&  \left[c_3 \hat{1}+ c_4\vec{L}_\text{p} \cdot \vec{S}+c_5(\vec{L}_\text{p} \cdot \vec{S})^2 \right] \nonumber \\
&\times \frac{9\delta(X)}{2X^4} |L_\text{p}=1,S=1 \rangle \langle L_\text{p}=1,S=1|  \,,
\label{eqn:second order}
\end{align}
which includes second order spin-orbit interaction. We obtain
\begin{align}
c_3&=\frac{-a(1,1,0,k)+3a(1,1,1,k)+a(1,1,2,k)}{3}\,, \nonumber\\
c_4&=\frac{-a(1,1,1,k)+a(1,1,2,k)}{2}\,, \nonumber\\
c_5&=\frac{2a(1,1,0,k)-3a(1,1,1,k)+a(1,1,2,k)}{6}\,.
\end{align}
The result implies that the $\vec{L}_\text{p} \cdot \vec{S}$ coupling vanishes, i.e. $c_4=c_5=0$, only if all scattering volumes are equal, which agrees with our expectation. Our analysis shows that including the second order spin-orbit interaction is particularly important for the description close to the $p$-wave shape resonance.

$\:$

$\:$

$\:$

$\:$

\begin{center}
	\LTcapwidth=\textwidth
	\begin{longtable*}[h]{c c c c || c c c c || c c c c c}
		\caption{Measured and calculated molecular energy level positions. The subscripts $\text{e}$ and $\text{t}$ denote experimental and theoretical results, respectively. $\Delta \nu_e$ and $\Delta \tilde{\nu}_e$ are measured resonance frequencies, while $\Delta \nu_\text{t}$ corresponds to computed term frequencies (referenced to the calculated $5S_{1/2}+16P_{1/2}$ dissociation threshold). The subscript $\text{s}$ indicates splittings between $|\Omega|$ states within individual multiplet structures and the subscript $\text{v}$ is used to mark vibrational splittings for a given $|\Omega|$ quantum number. Signal strengths of measured and calculated resonance lines are classified as weak ($w$) or strong ($s$). Not-observed lines are labeled with n.o. Values of $\Delta \nu_\text{e}$ and $\Delta \tilde{\nu}_e$ indicated by ($\ast$) characterize experimental signals which might come from different molecular states than considered here. These signals are not taken into account for Fig.$\:$\ref{fig5}. The resonance at $\Delta \tilde{\nu}_\text{e}=31.86\:\text{GHz}$ marked with ($\ast\ast$) is rather broad and expected to consist of an $N=5/2$ and an $N=3/2$ molecular line which cannot be resolved. Therefore, we give this frequency for the corresponding lines of the double as well as the triple line pattern.}	 \vspace{6.5pt} \label{table1} \\
		
		\toprule
		\toprule
		\multicolumn{4}{c}{Experiment ($F=1$)} & \multicolumn{4}{c}{Experiment ($F=2$)} & \multicolumn{5}{c}{Theory}  \\ \midrule
		\centering $\Delta\nu_\mathrm{e}$ & $\delta\nu_\text{s,e}$ & $\delta\nu_\text{v,e}$ & Signal strength & $\Delta\tilde{\nu}_\mathrm{e}$ & $\delta\tilde{\nu}_\text{s,e}$ & $\delta\tilde{\nu}_\text{v,e}$ & Signal strength & $|\Omega|$ & $\Delta\nu_\mathrm{t}$ & $\delta\nu_\text{s,t}$ & $\delta\nu_\text{v,t}$ & Signal strength  \\
		\centering (GHz)  & (GHz) & (GHz) & & (GHz) & (GHz) & (GHz) & & & (GHz) & (GHz) & (GHz) & \\ \midrule
		\multicolumn{12}{c}{Vibrational ladder of single lines ($N=1/2$, pure triplet)}                                                     \\ \midrule
		\endfirsthead
		
		\multicolumn{13}{c}%
		{{\tablename\ \thetable{}: Continuation}} \\
		\toprule
		\multicolumn{4}{c}{Experiment ($F=1$)} & \multicolumn{4}{c}{Experiment ($F=2$)} & \multicolumn{5}{c}{Theory} \\ \midrule
		\centering $\Delta\nu_\mathrm{e}$ & $\delta\nu_\text{s,e}$ & $\delta\nu_\text{v,e}$ & Signal strength & $\Delta\tilde{\nu}_\mathrm{e}$ & $\delta\tilde{\nu}_\text{s,e}$ & $\delta\tilde{\nu}_\text{v,e}$ & Signal strength & $|\Omega|$ & $\Delta\nu_\mathrm{t}$ & $\delta\nu_\text{s,t}$ & $\delta\nu_\text{v,t}$ & Signal strength \\
		\centering (GHz)  & (GHz) & (GHz) & & (GHz) & (GHz) & (GHz) & & & (GHz) & (GHz) & (GHz) & \\ \midrule
		\endhead
		
		\midrule \multicolumn{13}{r}{{Continued on next page}} \\ \midrule
		\endfoot
		
		\endlastfoot
		
		16.30 & & & $s$ &  &  &  &  & 0.5 & 14.36 & & & $s$ \\
		17.82 & & 1.52 & $w$ &  &  &  &  & 0.5 & 15.73 & & 1.37 & $w$ \\
		19.60 & & 1.78 & $s$ &  &  &  &  & 0.5 & 17.25 & & 1.52 & $s$ \\
		21.27 & & 1.67 & $w$ &  &  &  &  & 0.5 & 18.91 & & 1.66 & $w$ \\
		23.06 & & 1.79 & $s$ &  &  &  &  & 0.5 & 20.67 & & 1.76 & $s$ \\
		24.83 & & 1.77 & $w$ &  &  &  &  & 0.5 & 22.53 & & 1.86 & $w$ \\
		26.63 & & 1.80 & $s$ &  &  &  &  & 0.5 & 24.43 & & 1.90 & $s$ \\
		28.34 & & 1.71 & $w$ &  &  &  &  & 0.5 & 26.36 & & 1.93 & $w$ \\
		29.95 & & 1.61 & $s$ &  &  &  &  & 0.5 & 28.27 & & 1.91 & $s$ \\
		31.38$^{\ast}$ & & 1.43 & $w$ &  &  &  &  & 0.5 & 30.11 & & 1.84 & $w$ \\
		32.34$^{\ast}$ & & 0.96 & $s$ &  &  &  &  & 0.5 & 31.84 & & 1.73 & $s$ \\ \midrule
		\multicolumn{13}{c}{Double-line pattern ($N=3/2$, mixed singlet/triplet)}                                                     \\ \midrule
		18.68 & & & $s$ & 18.64 & & & $s$ & 1.5 & 16.86 & & & $s$ \\
		18.84 & 0.16 & & $s$ & 18.80 & 0.16 & & $s$ & 0.5 & 17.01 & 0.15 & & $s$ \\
		20.08 & & 1.40 & $w$ & 20.12 & & 1.48 & $w$ & 1.5 & 18.17 & & 1.31 & $w$ \\
		20.28 & 0.20 & 1.44 & $w$ & 20.32 & 0.20 & 1.52 & $w$ & 0.5 & 18.41 & 0.24 & 1.40 & $w$ \\
		21.74 & & 1.66 & $s$ & 21.76 & & 1.64 & $s$ & 1.5 & 19.63 & & 1.46 & $s$ \\
		21.96 & 0.22 & 1.68 & $s$ & 22.00 & 0.24 & 1.68 & $s$ & 0.5 & 19.95 & 0.32 & 1.54 & $s$ \\
		23.38 & & 1.64 & $w$ & n.o. & & & & 1.5 & 21.22 & & 1.59 & $w$ \\
		23.65 & 0.27 & 1.69 & $w$ & n.o. & & & & 0.5 & 21.61 & 0.39 & 1.66 & $w$ \\
		25.15 & & 1.77 & $s$ & 25.15 & & & $s$ & 1.5 & 22.93 & & 1.71 & $s$ \\
		25.43 & 0.28 & 1.78 & $s$ & 25.44 & 0.29 & & $s$ & 0.5 & 23.37 & 0.44 & 1.76 & $s$ \\
		26.87 & & 1.72 & $w$ & n.o. & & & & 1.5 & 24.72 & & 1.79 & $w$ \\
		27.16 & 0.29 & 1.73 & $w$ & n.o. & & & & 0.5 & 25.21 & 0.49 & 1.84 & $w$ \\
		28.62 & & 1.75 & $s$ & 28.65 & & & $s$ & 1.5 & 26.57 & & 1.85 & $s$ \\
		28.91 & 0.29 & 1.75 & $s$ & 28.93 & 0.28 & & $s$ & 0.5 & 27.10 & 0.53 & 1.89 & $s$ \\
		30.26 & & 1.64 & $w$ & n.o. & & & & 1.5 & 28.44 & & 1.87 & $w$ \\
		30.52 & 0.26 & 1.61 & $w$ & n.o. & & & & 0.5 & 28.98 & 0.54 & 1.88 & $w$ \\
		31.83 & & 1.57 & $s$ & 31.86$^{\ast\ast}$ & & & $s$ & 1.5 & 30.29 & & 1.85 & $s$ \\
		32.08 & 0.25 & 1.56 & $s$ & 32.10 & 0.24 & & $s$ & 0.5 & 30.83 & 0.54 & 1.85 & $s$ \\
		 &  &  &  &  &  & &  & 1.5 & 32.08 &  & 1.79 & $w$ \\
		 &  &  &  &  &  & &  & 0.5 & 32.58 & 0.50 & 1.75 & $w$ \\ \midrule
		\multicolumn{13}{c}{Triple-line pattern ($N=5/2$, pure triplet)}                                                     \\ \midrule
		& & & & 21.36$^{\ast}$ & & & $w$ & & & & & \\
		& & & & 21.43$^{\ast}$ & 0.07 & & $w$ & & & & & \\
		& & & & 21.54$^{\ast}$ & 0.11 & & $w$ & & & & & \\
		& & & & 23.08 & & 1.72 & $s$ & 0.5 & 21.12 & & & $s$ \\
		& & & & 23.15 & 0.07 & 1.72 & $s$ & 1.5 & 21.18 & 0.06 & & $s$ \\
		& & & & 23.25 & 0.10 & 1.71 & $s$ & 2.5 & 21.29 & 0.11 & & $s$ \\
		& & & & 24.59 & & 1.51 & $w$ & 0.5 & 22.47 & & 1.35 & $w$ \\
		& & & & 24.68 & 0.09 & 1.53 & $w$ & 1.5 & 22.56 & 0.09 & 1.38 & $w$ \\
		& & & & 24.81 & 0.13 & 1.56 & $w$ & 2.5 & 22.71 & 0.15 & 1.42 & $w$ \\
		& & & & 26.30 & & 1.71 & $s$ & 0.5 & 23.99 & & 1.52 & $s$ \\
		& & & & 26.39 & 0.09 & 1.71 & $s$ & 1.5 & 24.10 & 0.11 & 1.54 & $s$ \\
		& & & & 26.54 & 0.15 & 1.73 & $s$ & 2.5 & 24.28 & 0.18 & 1.57 & $s$ \\
		& & & & 28.01 & & 1.71 & $w$ & 0.5 & 25.65 & & 1.66 & $w$ \\
		& & & & 28.11 & 0.10 & 1.72 & $w$ & 1.5 & 25.78 & 0.13 & 1.68 & $w$ \\
		& & & & 28.27 & 0.16 & 1.73 & $w$ & 2.5 & 25.98 & 0.20 & 1.70 & $w$ \\
		& & & & 29.83 & & 1.82 & $s$ & 0.5 & 27.42 & & 1.77 & $s$ \\
		& & & & 29.94 & 0.11 & 1.83 & $s$ & 1.5 & 27.57 & 0.15 & 1.79 & $s$ \\
		& & & & 30.11 & 0.17 & 1.84 & $s$ & 2.5 & 27.79 & 0.22 & 1.81 & $s$ \\
		& & & & 31.61 & & 1.78 & $w$ & 0.5 & 29.28 & & 1.86 & $w$ \\
		& & & & 31.72 & 0.11 & 1.78 & $w$ & 1.5 & 29.44 & 0.16 & 1.87 & $w$ \\
		& & & & 31.86$^{\ast\ast}$ & 0.14 & 1.75 & $w$ & 2.5 & 29.67 & 0.23 & 1.88 & $w$ \\
		& & & & 33.42 & & 1.81 & $s$ & 0.5 & 31.19 & & 1.91 & $s$ \\
		& & & & 33.53 & 0.11 & 1.81 & $s$ & 1.5 & 31.36 & 0.17 & 1.92 & $s$ \\
		& & & & 33.72 & 0.19 & 1.86 & $s$ & 2.5 & 31.59 & 0.23 & 1.92 & $s$ \\
		& & & & n.o. & & & & 0.5 & 33.11 & & 1.92 & $w$ \\
		& & & & n.o. & & & & 1.5 & 33.28 & 0.17 & 1.92 & $w$ \\
		& & & & n.o. & & & & 2.5 & 33.51 & 0.23 & 1.92 & $w$ \\
		& & & & 36.74 & & & $s$ & 0.5 & 35.00 & & 1.89 & $s$ \\
		& & & & 36.85 & 0.11 & & $s$ & 1.5 & 35.16 & 0.16 & 1.88 & $s$ \\
		& & & & 37.00 & 0.15 & & $s$ & 2.5 & 35.37 & 0.21 & 1.86 & $s$ \\
		& & & & & & & & 0.5 & 36.83 & & 1.83 & $w$ \\
		& & & & & & & & 1.5 & 36.96 & 0.13 & 1.80 & $w$ \\
		& & & & & & & & 2.5 & 37.14 & 0.18 & 1.77 & $w$ \\ \bottomrule
		\bottomrule
	\end{longtable*}
\end{center}


\begin{thebibliography}{30}%
\makeatletter
\providecommand \@ifxundefined [1]{%
 \@ifx{#1\undefined}
}%
\providecommand \@ifnum [1]{%
 \ifnum #1\expandafter \@firstoftwo
 \else \expandafter \@secondoftwo
 \fi
}%
\providecommand \@ifx [1]{%
 \ifx #1\expandafter \@firstoftwo
 \else \expandafter \@secondoftwo
 \fi
}%
\providecommand \natexlab [1]{#1}%
\providecommand \enquote  [1]{``#1''}%
\providecommand \bibnamefont  [1]{#1}%
\providecommand \bibfnamefont [1]{#1}%
\providecommand \citenamefont [1]{#1}%
\providecommand \href@noop [0]{\@secondoftwo}%
\providecommand \href [0]{\begingroup \@sanitize@url \@href}%
\providecommand \@href[1]{\@@startlink{#1}\@@href}%
\providecommand \@@href[1]{\endgroup#1\@@endlink}%
\providecommand \@sanitize@url [0]{\catcode `\\12\catcode `\$12\catcode
  `\&12\catcode `\#12\catcode `\^12\catcode `\_12\catcode `\%12\relax}%
\providecommand \@@startlink[1]{}%
\providecommand \@@endlink[0]{}%
\providecommand \url  [0]{\begingroup\@sanitize@url \@url }%
\providecommand \@url [1]{\endgroup\@href {#1}{\urlprefix }}%
\providecommand \urlprefix  [0]{URL }%
\providecommand \Eprint [0]{\href }%
\providecommand \doibase [0]{http://dx.doi.org/}%
\providecommand \selectlanguage [0]{\@gobble}%
\providecommand \bibinfo  [0]{\@secondoftwo}%
\providecommand \bibfield  [0]{\@secondoftwo}%
\providecommand \translation [1]{[#1]}%
\providecommand \BibitemOpen [0]{}%
\providecommand \bibitemStop [0]{}%
\providecommand \bibitemNoStop [0]{.\EOS\space}%
\providecommand \EOS [0]{\spacefactor3000\relax}%
\providecommand \BibitemShut  [1]{\csname bibitem#1\endcsname}%
\let\auto@bib@innerbib\@empty

\bibitem{Greene2000}
\bibinfo{author}{C.~H. Greene}, \bibinfo{author}{A.~S. Dickinson}, and \bibinfo{author}{H.~R. Sadeghpour},
\href {\doibase  10.1103/PhysRevLett.85.2458}{\bibinfo{journal}{Phys. Rev. Lett.}
	\textbf{\bibinfo{volume}{85}}, \bibinfo{pages}{2458}
	(\bibinfo{year}{2000})}.

\bibitem{Bendkowsky2009}
\bibinfo{author}{V. Bendkowsky}, \bibinfo{author}{B. Butscher}, \bibinfo{author}{J. Nipper}, \bibinfo{author}{J.~P. Shaffer}, \bibinfo{author}{R. L\"{o}w}, and \bibinfo{author}{T. Pfau},
\href {\doibase 10.1038/nature07945}{\bibinfo{journal}{Nature}
	\textbf{\bibinfo{volume}{458}}, \bibinfo{pages}{1005}
	(\bibinfo{year}{2009})}.

\bibitem{Shaffer2018}
\bibinfo{author}{J.~P. Shaffer}, \bibinfo{author}{S.~T. Rittenhouse}, and \bibinfo{author}{H.~R. Sadeghpour},
\href {\doibase  10.1038/s41467-018-04135-6}{\bibinfo{journal}{Nat. Commun.}
	\textbf{\bibinfo{volume}{9}}, \bibinfo{pages}{1965}
	(\bibinfo{year}{2018})}.

\bibitem{Fey2019}
\bibinfo{author}{C. Fey}, \bibinfo{author}{F. Hummel}, and \bibinfo{author}{P. Schmelcher},
\href {\doibase 10.1080/00268976.2019.1679401}{\bibinfo{journal}{Mol. Phys.}
	(\bibinfo{year}{2019})}, doi:10.1080/00268976.2019.1679401.

\bibitem{Eiles2019}
\bibinfo{author}{M.~T. Eiles},
\href {\doibase 10.1088/1361-6455/ab19ca}{\bibinfo{journal}{J. Phys. B: At. Mol. Opt. Phys.}
	\textbf{\bibinfo{volume}{52}}, \bibinfo{pages}{113001}
	(\bibinfo{year}{2019})}.

\bibitem{Khuskivadze2002}
\bibinfo{author}{A.~A. Khuskivadze}, \bibinfo{author}{M.~I. Chibisov}, and \bibinfo{author}{I.~I. Fabrikant},
\href {\doibase 10.1103/PhysRevA.66.042709}{\bibinfo{journal}{Phys. Rev. A}
	\textbf{\bibinfo{volume}{66}}, \bibinfo{pages}{042709}
	(\bibinfo{year}{2002})}.

\bibitem{Eiles2017}
\bibinfo{author}{M.~T. Eiles} and \bibinfo{author}{C.~H. Greene},
\href {\doibase  10.1103/PhysRevA.95.042515}{\bibinfo{journal}{Phys. Rev. A}
	\textbf{\bibinfo{volume}{95}}, \bibinfo{pages}{042515}
	(\bibinfo{year}{2017})}.

\bibitem{Markson2016}
\bibinfo{author}{S. Markson}, \bibinfo{author}{S.~T. Rittenhouse}, \bibinfo{author}{R. Schmidt}, \bibinfo{author}{J.~P. Shaffer}, and \bibinfo{author}{H.~R. Sadeghpour},
\href {\doibase 10.1002/cphc.201600932}{\bibinfo{journal}{Chem. Phys. Chem.}
	\textbf{\bibinfo{volume}{17}}, \bibinfo{pages}{3683}
	(\bibinfo{year}{2016})}.

\bibitem{Thomas2018}
\bibinfo{author}{O. Thomas}, \bibinfo{author}{C. Lippe}, \bibinfo{author}{T. Eichert}, and \bibinfo{author}{H. Ott},
\href {\doibase  10.1038/s41467-018-04684-w}{\bibinfo{journal}{Nat. Commun.}
	\textbf{\bibinfo{volume}{9}}, \bibinfo{pages}{2238}
	(\bibinfo{year}{2018})}.

\bibitem{Li2011}
\bibinfo{author}{W. Li}, \bibinfo{author}{T. Pohl}, \bibinfo{author}{J.~M. Rost}, \bibinfo{author}{S.~T. Rittenhouse}, \bibinfo{author}{H.~R. Sadeghpour}, \bibinfo{author}{J. Nipper}, \bibinfo{author}{B. Butscher}, \bibinfo{author}{J.~B. Balewski}, \bibinfo{author}{V. Bendkowsky}, \bibinfo{author}{R. L\"{o}w}, and \bibinfo{author}{T. Pfau},
\href {\doibase 10.1126/science.1211255}{\bibinfo{journal}{Science}
	\textbf{\bibinfo{volume}{334}}, \bibinfo{pages}{1110}
	(\bibinfo{year}{2011})}.

\bibitem{Tallant2012}
\bibinfo{author}{J. Tallant}, \bibinfo{author}{S.~T. Rittenhouse}, \bibinfo{author}{D. Booth}, \bibinfo{author}{H.~R. Sadeghpour}, and \bibinfo{author}{J.~P. Shaffer},
\href {\doibase 10.1103/PhysRevLett.109.173202}{\bibinfo{journal}{Phys. Rev. Lett.}
	\textbf{\bibinfo{volume}{109}}, \bibinfo{pages}{173202}
	(\bibinfo{year}{2012})}.

\bibitem{Booth2015}
\bibinfo{author}{D. Booth}, \bibinfo{author}{S.~T. Rittenhouse}, \bibinfo{author}{J. Yang}, \bibinfo{author}{H.~R. Sadeghpour}, and \bibinfo{author}{J.~P. Shaffer},
\href {\doibase 10.1126/science.1260722}{\bibinfo{journal}{Science}
	\textbf{\bibinfo{volume}{348}}, \bibinfo{pages}{99}
	(\bibinfo{year}{2015})}.

\bibitem{Niederpruem2016a}
\bibinfo{author}{T. Niederpr\"{u}m}, \bibinfo{author}{O. Thomas}, \bibinfo{author}{T. Eichert}, \bibinfo{author}{C. Lippe}, \bibinfo{author}{J. P\'{e}rez-R\'{i}os}, \bibinfo{author}{C.~H. Greene}, and \bibinfo{author}{H. Ott},
\href {\doibase  10.1038/ncomms12820}{\bibinfo{journal}{Nat. Commun.}
	\textbf{\bibinfo{volume}{7}}, \bibinfo{pages}{12820}
	(\bibinfo{year}{2016})}.

\bibitem{Krupp2014}
\bibinfo{author}{A.~T. Krupp}, \bibinfo{author}{A. Gaj}, \bibinfo{author}{J.~B. Balewski}, \bibinfo{author}{P. Ilzh\"{o}fer}, \bibinfo{author}{S. Hofferberth}, \bibinfo{author}{R. L\"{o}w}, \bibinfo{author}{T. Pfau}, \bibinfo{author}{M. Kurz}, and \bibinfo{author}{P. Schmelcher},
\href {\doibase 10.1103/PhysRevLett.112.143008}{\bibinfo{journal}{Phys. Rev. Lett.}
	\textbf{\bibinfo{volume}{112}}, \bibinfo{pages}{143008}
	(\bibinfo{year}{2014})}.

\bibitem{DeSalvo2015}
\bibinfo{author}{B.~J. DeSalvo}, \bibinfo{author}{J.~A. Aman}, \bibinfo{author}{F.~B. Dunning}, \bibinfo{author}{T.~C. Killian}, \bibinfo{author}{H.~R. Sadeghpour}, \bibinfo{author}{S. Yoshida}, and \bibinfo{author}{J. Burgd\"{o}rfer},
\href {\doibase 10.1103/PhysRevA.92.031403}{\bibinfo{journal}{Phys. Rev. A}
	\textbf{\bibinfo{volume}{92}}, \bibinfo{pages}{031403(R)}
	(\bibinfo{year}{2015})}.

\bibitem{Bellos2013}
\bibinfo{author}{M.~A. Bellos}, \bibinfo{author}{R. Carollo}, \bibinfo{author}{J. Banerjee}, \bibinfo{author}{E.~E. Eyler}, \bibinfo{author}{P.~L. Gould}, and \bibinfo{author}{W.~C. Stwalley},
\href {\doibase 10.1103/PhysRevLett.111.053001}{\bibinfo{journal}{Phys. Rev. Lett.}
	\textbf{\bibinfo{volume}{111}}, \bibinfo{pages}{053001}
	(\bibinfo{year}{2013})}.

\bibitem{Kleinbach2017}
\bibinfo{author}{K.~S. Kleinbach},  \bibinfo{author}{F. Meinert}, \bibinfo{author}{F. Engel}, \bibinfo{author}{W.~J. Kwon}, \bibinfo{author}{R. L\"{o}w}, \bibinfo{author}{T. Pfau}, and \bibinfo{author}{G. Raithel},
\href {\doibase 10.1103/PhysRevLett.118.223001}{\bibinfo{journal}{Phys. Rev. Lett.} \textbf{\bibinfo{volume}{118}}, \bibinfo{pages}{223001} (\bibinfo{year}{2017})}.

\bibitem{Bendkowsky2010}
\bibinfo{author}{V. Bendkowsky}, \bibinfo{author}{B. Butscher}, \bibinfo{author}{J. Nipper}, \bibinfo{author}{J.~P. Balewski}, \bibinfo{author}{J.~P. Shaffer}, \bibinfo{author}{R. L\"{o}w}, \bibinfo{author}{T. Pfau}, \bibinfo{author}{W. Li}, \bibinfo{author}{J. Stanojevic}, \bibinfo{author}{T. Pohl}, and \bibinfo{author}{J.~M. Rost},
\href {\doibase 10.1103/PhysRevLett.105.163201}{\bibinfo{journal}{Phys. Rev. Lett.}
	\textbf{\bibinfo{volume}{105}}, \bibinfo{pages}{163201}
	(\bibinfo{year}{2010})}.

\bibitem{Boettcher2016}
\bibinfo{author}{F. B\"{o}ttcher}, \bibinfo{author}{A. Gaj}, \bibinfo{author}{K.~M. Westphal}, \bibinfo{author}{M. Schlagm\"{u}ller}, \bibinfo{author}{K.~S. Kleinbach}, \bibinfo{author}{R. L\"{o}w}, \bibinfo{author}{T. Cubel Liebisch}, \bibinfo{author}{T. Pfau}, and \bibinfo{author}{S. Hofferberth},
\href {\doibase  10.1103/PhysRevA.93.032512}{\bibinfo{journal}{Phys. Rev. A}
	\textbf{\bibinfo{volume}{93}}, \bibinfo{pages}{032512}
	(\bibinfo{year}{2016})}.

\bibitem{Sassmannshausen2015}
\bibinfo{author}{H. Sa{\ss}mannshausen}, \bibinfo{author}{F. Merkt}, and \bibinfo{author}{J. Deiglmayr},
\href {\doibase 10.1103/PhysRevLett.114.133201}{\bibinfo{journal}{Phys. Rev. Lett.}
	\textbf{\bibinfo{volume}{114}}, \bibinfo{pages}{133201}
	(\bibinfo{year}{2015})}.

\bibitem{MacLennan2018}
\bibinfo{author}{J.~L. MacLennan}, \bibinfo{author}{Y.-J. Chen}, and \bibinfo{author}{G. Raithel},
\href {\doibase  10.1103/PhysRevA.99.033407}{\bibinfo{journal}{Phys. Rev. A}
	\textbf{\bibinfo{volume}{99}}, \bibinfo{pages}{033407}
	(\bibinfo{year}{2019})}.

\bibitem{Niederpruem2016b}
\bibinfo{author}{T. Niederpr\"{u}m}, \bibinfo{author}{O. Thomas}, \bibinfo{author}{T. Eichert}, and \bibinfo{author}{H. Ott},
\href {\doibase  10.1103/PhysRevLett.117.123002}{\bibinfo{journal}{Phys. Rev. Lett.}
	\textbf{\bibinfo{volume}{117}}, \bibinfo{pages}{123002}
	(\bibinfo{year}{2016})}.

\bibitem{Anderson2014a}
\bibinfo{author}{D.~A. Anderson}, \bibinfo{author}{S.~A. Miller}, and \bibinfo{author}{G. Raithel},
\href {\doibase 10.1103/PhysRevLett.112.163201}{\bibinfo{journal}{Phys. Rev. Lett.}
	\textbf{\bibinfo{volume}{112}}, \bibinfo{pages}{163201}
	(\bibinfo{year}{2014})}.

\bibitem{Engel2019}
\bibinfo{author}{F. Engel}, \bibinfo{author}{T. Dieterle}, \bibinfo{author}{F. Hummel}, \bibinfo{author}{C. Fey}, \bibinfo{author}{P. Schmelcher}, \bibinfo{author}{R. L\"{o}w}, \bibinfo{author}{T. Pfau}, and \bibinfo{author}{F. Meinert},
\href {\doibase	10.1103/PhysRevLett.123.073003}{\bibinfo{journal}{Phys. Rev. Lett.}
	\textbf{\bibinfo{volume}{123}}, \bibinfo{pages}{073003}
	(\bibinfo{year}{2019})}.

\bibitem{Hummel2018}
\bibinfo{author}{F. Hummel}, \bibinfo{author}{C. Fey}, and \bibinfo{author}{P. Schmelcher}, \href {\doibase 10.1103/PhysRevA.99.023401}{\bibinfo{journal}{Phys. Rev. A},
	\textbf{\bibinfo{volume}{99}}, \bibinfo{pages}{023401} (\bibinfo{year}{2019})}.

\bibitem{Anderson2014b}
\bibinfo{author}{D.~A. Anderson}, \bibinfo{author}{S.~A. Miller}, and \bibinfo{author}{G. Raithel},
\href {\doibase 10.1103/PhysRevA.90.062518}{\bibinfo{journal}{Phys. Rev. A}
	\textbf{\bibinfo{volume}{90}}, \bibinfo{pages}{062518}
	(\bibinfo{year}{2014})}.

\bibitem{Hamilton2002}
\bibinfo{author}{E.~L. Hamilton}, \bibinfo{author}{C.~H. Greene}, and \bibinfo{author}{H.~R. Sadeghpour},
\href {\doibase  10.1088/0953-4075/35/10/102}{\bibinfo{journal}{J. Phys. B: At. Mol. Opt. Phys.}
	\textbf{\bibinfo{volume}{35}}, \bibinfo{pages}{L199}
	(\bibinfo{year}{2002})}.

\bibitem{Chibisov2002}
\bibinfo{author}{M.~I. Chibisov}, \bibinfo{author}{A.~A. Khuskivadze}, and \bibinfo{author}{I.~I. Fabrikant},
\href {\doibase  10.1088/0953-4075/35/10/101}{\bibinfo{journal}{J. Phys. B: At. Mol. Opt. Phys.}
	\textbf{\bibinfo{volume}{35}}, \bibinfo{pages}{L193}
	(\bibinfo{year}{2002})}.

\bibitem{noteEavg}
$E_r^{\text{avg}}$ represents an average value which follows the rule $E_r^\text{avg} = \left[ \sum_{J=0}^2 (2J+1) E_r^J \right]/ \left[\sum_{J=0}^2 (2J+1) \right]$. Here, $E_r^J$ are the characteristic energies where the $p$-wave shape resonance occurs for the three individual channels $J=\{0,\,1,\,2\}$ (see also the discussion in Sec.$\:$\ref{subsec:LSmodel}).

\bibitem{Schmid2012}
\bibinfo{author}{S. Schmid},  \bibinfo{author}{A. H\"{a}rter}, \bibinfo{author}{A. Frisch}, \bibinfo{author}{S. Hoinka}, and \bibinfo{author}{J. Hecker Denschlag},
\href {\doibase 10.1063/1.4718356}{\bibinfo{journal}{Rev. Sci. Instrum.} \textbf{\bibinfo{volume}{83}}, \bibinfo{pages}{053108} (\bibinfo{year}{2012})}.

\bibitem{Haerter2013b}
\bibinfo{author}{A. H\"{a}rter}, \bibinfo{author}{A. Kr\"{u}kow}, \bibinfo{author}{A. Brunner}, and
\bibinfo{author}{J. Hecker Denschlag},
\href {\doibase 10.1063/1.4809578}{\bibinfo{journal}{Appl. Phys. Lett.}
	\textbf{\bibinfo{volume}{102}}, \bibinfo{pages}{221115}
	(\bibinfo{year}{2013})}.

\bibitem{Wolf2017}
\bibinfo{author}{J. Wolf}, \bibinfo{author}{M. Dei{\ss}}, \bibinfo{author}{A. Kr\"{u}kow}, \bibinfo{author}{E. Tiemann}, \bibinfo{author}{B.~P. Ruzic}, \bibinfo{author}{Y. Wang}, \bibinfo{author}{J.~P. D'Incao}, \bibinfo{author}{P.~S. Julienne}, and \bibinfo{author}{J. Hecker Denschlag},
\href {\doibase  10.1126/science.aan8721}{\bibinfo{journal}{Science}
	\textbf{\bibinfo{volume}{358}}, \bibinfo{pages}{921}
	(\bibinfo{year}{2017})}.


\bibitem{Niederpruem2015}
\bibinfo{author}{T. Niederpr\"{u}m}, \bibinfo{author}{O. Thomas}, \bibinfo{author}{T. Manthey}, \bibinfo{author}{T.~M. Weber}, and \bibinfo{author}{H. Ott},
\href {\doibase  10.1103/PhysRevLett.115.013003}{\bibinfo{journal}{Phys. Rev. Lett.}
	\textbf{\bibinfo{volume}{115}}, \bibinfo{pages}{013003}
	(\bibinfo{year}{2015})}.

\bibitem{Schlagmueller2016}
\bibinfo{author}{M. Schlagm\"{u}ller}, \bibinfo{author}{T. Cubel Liebisch}, \bibinfo{author}{F. Engel}, \bibinfo{author}{K.~S. Kleinbach}, \bibinfo{author}{F. B\"{o}ttcher}, \bibinfo{author}{U. Hermann}, \bibinfo{author}{K.~M. Westphal}, \bibinfo{author}{A. Gaj}, \bibinfo{author}{R. L\"{o}w}, \bibinfo{author}{S. Hofferberth}, \bibinfo{author}{T. Pfau}, \bibinfo{author}{J. P\'{e}rez-R\'{i}os}, and \bibinfo{author}{C.~H. Greene},
\href {\doibase  10.1103/PhysRevX.6.031020}{\bibinfo{journal}{Phys. Rev. X}
	\textbf{\bibinfo{volume}{6}}, \bibinfo{pages}{031020}
	(\bibinfo{year}{2016})}.

\bibitem{Bize1999}
\bibinfo{author}{S. Bize}, \bibinfo{author}{Y. Sortais}, \bibinfo{author}{M.~S. Santos}, \bibinfo{author}{C. Mandache}, \bibinfo{author}{A. Clairon}, and
\bibinfo{author}{C. Salomon},
\href {\doibase 10.1209/epl/i1999-00203-9}{\bibinfo{journal}{Europhys. Lett.}
	\textbf{\bibinfo{volume}{45}}, \bibinfo{pages}{558}
	(\bibinfo{year}{1999})}.

\bibitem{Arimondo1977}
\bibinfo{author}{E. Arimondo}, \bibinfo{author}{M. Inguscio}, and \bibinfo{author}{P. Violino},
\href {\doibase 10.1103/RevModPhys.49.31}{\bibinfo{journal}{Rev. Mod. Phys.}
	\textbf{\bibinfo{volume}{49}}, \bibinfo{pages}{31} (\bibinfo{year}{1977})}.

\bibitem{Fey2015}
\bibinfo{author}{C. Fey}, \bibinfo{author}{M. Kurz}, \bibinfo{author}{P. Schmelcher}, \bibinfo{author}{S.~T. Rittenhouse}, and \bibinfo{author}{H.~R. Sadeghpour},
\href {\doibase 10.1088/1367-2630/17/5/055010}{\bibinfo{journal}{New J. Phys.} \textbf{\bibinfo{volume}{17}}, \bibinfo{pages}{055010} (\bibinfo{year}{2015})}.

\bibitem{Li2003}
\bibinfo{author}{W. Li}, \bibinfo{author}{I. Mourachko}, \bibinfo{author}{M.~W. Noel}, and \bibinfo{author}{T.~F. Gallagher},
\href {\doibase 10.1103/PhysRevA.67.052502}{\bibinfo{journal}{Phys. Rev. A},
	\textbf{\bibinfo{volume}{67}}, \bibinfo{pages}{052502} (\bibinfo{year}{2003})}.

\bibitem{Han2006}
\bibinfo{author}{J. Han}, \bibinfo{author}{Y. Jamil}, \bibinfo{author}{D.~V.~L. Norum}, \bibinfo{author}{P.~J. Tanner}, and \bibinfo{author}{T.~F. Gallagher},
\href {\doibase 10.1103/PhysRevA.74.054502}{\bibinfo{journal}{Phys. Rev. A},
	\textbf{\bibinfo{volume}{74}}, \bibinfo{pages}{054502} (\bibinfo{year}{2006})}.


\bibitem{Hummel2017}
\bibinfo{author}{F. Hummel}, \bibinfo{author}{C. Fey}, and \bibinfo{author}{P. Schmelcher},
\href {\doibase 10.1103/PhysRevA.97.043422}{\bibinfo{journal}{Phys. Rev. A},
	\textbf{\bibinfo{volume}{97}}, \bibinfo{pages}{043422} (\bibinfo{year}{2018})}.

\bibitem{note}
The operator $\overleftarrow{\nabla}_{\vec{r}} \cdot \delta(\vec{R}-\vec{r}) \overrightarrow{\nabla}_{\vec{r}}$ is a shorthand for
$ \sum_{i,j}  | \psi_i  \rangle  \langle \psi_j | \,  \int d^3r \delta(\vec{R}-\vec{r}) \left(\overrightarrow{\nabla}_{\vec{r}}\psi^*_i(\vec{r})\right) \cdot \left(\overrightarrow{\nabla}_{\vec{r}} \psi_j(\vec{r})\right)  $, where the $ \{ | \psi_i \rangle \} $ form an orthonormal basis of the Hilbert space and $ \psi_i(\vec{r}) \equiv \langle \vec{r}  | \psi_i \rangle   $.




\bibitem{Koeppel1984}
\bibinfo{author}{H. K\"{o}ppel}, \bibinfo{author}{W. Domcke}, and \bibinfo{author}{L.~S. Cederbaum},
\href {\doibase 10.1002/9780470142813.ch2}{\bibinfo{journal}{Adv. Chem. Phys.} \textbf{\bibinfo{volume}{57}}, \bibinfo{pages}{59} (\bibinfo{year}{1984})}.

\bibitem{Worth2004}
\bibinfo{author}{G.~A. Worth} and \bibinfo{author}{L.~S. Cederbaum},
\href {\doibase 10.1146/annurev.physchem.55.091602.094335}{\bibinfo{journal}{Annu. Rev. Phys. Chem.} \textbf{\bibinfo{volume}{55}}, \bibinfo{pages}{127} (\bibinfo{year}{2004})}.

\bibitem{Tarana2016}
\bibinfo{author}{M. Tarana} and \bibinfo{author}{R. \v{C}ur\'{i}k},
\href {\doibase 10.1103/PhysRevA.93.012515}{\bibinfo{journal}{Phys. Rev. A} \textbf{\bibinfo{volume}{93}}, \bibinfo{pages}{012515} (\bibinfo{year}{2016})}.

\bibitem{Fey2016}
\bibinfo{author}{C. Fey}, \bibinfo{author}{M. Kurz}, and \bibinfo{author}{P. Schmelcher},
\href {\doibase 10.1103/PhysRevA.94.012516}{\bibinfo{journal}{Phys. Rev. A}
	\textbf{\bibinfo{volume}{94}}, \bibinfo{pages}{012516}
	(\bibinfo{year}{2016})}.

\bibitem{Schlagmueller2016b}
\bibinfo{author}{M. Schlagm\"{u}ller}, \bibinfo{author}{T. Cubel Liebisch}, \bibinfo{author}{H. Nguyen}, \bibinfo{author}{G. Lochead}, \bibinfo{author}{F. Engel}, \bibinfo{author}{F. B\"{o}ttcher}, \bibinfo{author}{K.~M. Westphal}, \bibinfo{author}{K.~S. Kleinbach}, \bibinfo{author}{R. L\"{o}w}, \bibinfo{author}{S. Hofferberth}, \bibinfo{author}{T. Pfau}, \bibinfo{author}{J. P\'{e}rez-R\'{i}os}, and \bibinfo{author}{C.~H. Greene},
\href {\doibase 10.1103/PhysRevLett.116.053001}{\bibinfo{journal}{Phys. Rev. Lett.}
	\textbf{\bibinfo{volume}{116}}, \bibinfo{pages}{053001}
	(\bibinfo{year}{2016})}.

\bibitem{Eiles2016}
\bibinfo{author}{M.~T. Eiles}, \bibinfo{author}{J. P\'{e}rez-R\'{i}os}, \bibinfo{author}{F. Robicheaux}, and \bibinfo{author}{C.~H. Greene},
\href {\doibase 10.1088/0953-4075/49/11/114005}{\bibinfo{journal}{J. Phys. B: At. Mol. Opt. Phys.}
	\textbf{\bibinfo{volume}{49}}, \bibinfo{pages}{114005}
	(\bibinfo{year}{2016})}.

\bibitem{Schmidt2016}
\bibinfo{author}{R. Schmidt}, \bibinfo{author}{H.~R. Sadeghpour}, and \bibinfo{author}{E. Demler},
\href {\doibase 10.1103/PhysRevLett.116.105302}{\bibinfo{journal}{Phys. Rev. Lett.}
	\textbf{\bibinfo{volume}{116}}, \bibinfo{pages}{105302}
	(\bibinfo{year}{2016})}.

\bibitem{Ashida2019}
\bibinfo{author}{Y. Ashida}, \bibinfo{author}{T. Shi}, \bibinfo{author}{R. Schmidt}, \bibinfo{author}{H.~R. Sadeghpour}, \bibinfo{author}{J.~I. Cirac}, and \bibinfo{author}{E. Demler},
\href {\doibase 10.1103/PhysRevLett.123.183001}{\bibinfo{journal}{Phys. Rev. Lett.}
	\textbf{\bibinfo{volume}{123}}, \bibinfo{pages}{183001}
	(\bibinfo{year}{2019})}.

\bibitem{Haze2019}
\bibinfo{author}{S. Haze}, \bibinfo{author}{J. Wolf}, \bibinfo{author}{M. Dei{\ss}}, \bibinfo{author}{L. Wang}, \bibinfo{author}{G. Raithel}, and
\bibinfo{author}{J. Hecker Denschlag},
{\bibinfo{journal}{arXiv:1901.11069}
	(\bibinfo{year}{2019})}.

\bibitem{Ewald2018}
\bibinfo{author}{N.~V. Ewald}, \bibinfo{author}{T. Feldker}, \bibinfo{author}{H. Hirzler}, \bibinfo{author}{H. F\"{u}rst}, and
\bibinfo{author}{R. Gerritsma},
\href {\doibase 10.1103/PhysRevLett.122.253401}{\bibinfo{journal}{Phys. Rev. Lett.} \textbf{\bibinfo{volume}{122}}, \bibinfo{pages}{253401} (\bibinfo{year}{2019})}.

\bibitem{Engel2018}
\bibinfo{author}{F. Engel},  \bibinfo{author}{T. Dieterle}, \bibinfo{author}{T. Schmid}, \bibinfo{author}{C. Tomschitz}, \bibinfo{author}{C. Veit}, \bibinfo{author}{N. Zuber}, \bibinfo{author}{R. L\"{o}w}, \bibinfo{author}{T. Pfau}, and \bibinfo{author}{F. Meinert},
\href {\doibase 10.1103/PhysRevLett.121.193401}{\bibinfo{journal}{Phys. Rev. Lett.} \textbf{\bibinfo{volume}{121}}, \bibinfo{pages}{193401} (\bibinfo{year}{2018})}.

\bibitem{Theodosiou1984}
\bibinfo{author}{C.~E. Theodosiou},
\href {\doibase 10.1103/PhysRevA.30.2881}{\bibinfo{journal}{Phys. Rev. A} \textbf{\bibinfo{volume}{30}}, \bibinfo{pages}{2881} (\bibinfo{year}{1984})}.


\end{thebibliography}
\end{document}